\documentclass[letterpaper, twocolumn, superscriptaddress, showpacs, preprintnumbers, amsmath, amssymb,longbibliography, nofootinbib]{revtex4-1}
\usepackage[pdftex]{graphicx}
\usepackage{caption}
\usepackage{subcaption}
\usepackage{dcolumn}
\usepackage{bm}
\usepackage{color}
\usepackage{hyperref}
\usepackage{adjustbox}
\usepackage{color}
\usepackage{float}
\usepackage{framed}
\usepackage{esvect}
\usepackage{amsfonts,amssymb,amsmath,hyperref,hypcap,enumerate}
\usepackage{wasysym}
\usepackage{slashed}
\usepackage{tikz}
\usepackage{amsmath,amssymb,bm,graphicx,dsfont}
\usepackage[latin9]{inputenc}
\usepackage{amsmath}
\usepackage{slashed}
\usepackage{dsfont}
\usepackage{amstext}
\usepackage{amssymb}
\usepackage{amsbsy}
\usepackage{amsthm}
\usepackage{epsfig}
\usepackage{tikz}
\usepackage{bbm}
\usepackage{hyperref}
\usepackage{color}
\usepackage{multirow}
\newcommand{\up}{\uparrow}
\newcommand{\down}{\downarrow}

\newcommand{\T}{\mathcal{T}}

\newcommand{\be}{\begin{equation}}
\newcommand{\ba}{\begin{align}}
\newcommand{\ee}{\end{equation}}
\newcommand{\bea}{\begin{eqnarray}}
\newcommand{\eea}{\end{eqnarray}}
\newcommand{\beq}{\begin{equation}}
\newcommand{\eeq}{\end{equation}}
\newcommand{\beqn}{\begin{eqnarray}}
\newcommand{\eeqn}{\end{eqnarray}}

\newcommand{\RR}{{\mathcal R}}
\newcommand{\CC}{\mathcal{C}}

\newcommand{\ket}{\rangle}
\newcommand{\bra}{\langle}

\renewcommand{\vec}[1]{{\bf #1}}
\renewcommand{\hat}[1]{{\widehat #1}}

\def\nn{\nonumber\\}

\begin{document}
\title{ From spinon band topology to the symmetry quantum numbers of monopoles in Dirac spin liquids }
\author{Xue-Yang Song}
\affiliation{Department of Physics,  Harvard University,
Cambridge, MA 02138, USA}
\author{Yin-Chen He}
\affiliation{Perimeter Institute for Theoretical Physics, Waterloo, ON N2L 2Y5, Canada}
\affiliation{Department of Physics,  Harvard University,
Cambridge, MA 02138, USA}
\author{Ashvin Vishwanath}
\affiliation{Department of Physics,  Harvard University,
Cambridge, MA 02138, USA}
\author{Chong Wang}
\affiliation{Perimeter Institute for Theoretical Physics, Waterloo, ON N2L 2Y5, Canada}
\affiliation{Department of Physics,  Harvard University,
Cambridge, MA 02138, USA}
\date{\today}

\begin{abstract}
The interplay of symmetry and topology has been at the forefront of recent progress in quantum matter. Here we uncover an unexpected connection between band topology and the description of competing orders in a quantum magnet. Specifically we show that aspects of band topology protected by crystalline symmetries determine key properties of the Dirac spin liquid (DSL) which can be defined on the honeycomb, square, triangular and kagom\'e lattices. At low energies, the DSL on all these lattices is described by an emergent Quantum Electrodynamics (QED$_3$) with $N_f=4$ flavors of Dirac fermions coupled to a $U(1)$ gauge field. However the symmetry properties of the magnetic monopoles, an important class of critical degrees of freedom, behave very differently on different lattices. 
In particular, we show that the lattice momentum and angular momentum of monopoles can be determined from the charge (or Wannier) centers of the corresponding spinon insulator. We also show that for DSLs on bipartite lattices, there always exists a monopole that transforms trivially under all microscopic symmetries owing to the existence of a parent SU(2) gauge theory. We connect our results to generalized Lieb-Schultz-Mattis theorems and also derive the time-reversal and reflection properties of monopoles. Our results indicate that recent insights into free fermion band topology can also guide the description of strongly correlated quantum matter. 

\end{abstract}

\date{\today}

\maketitle

\maketitle

\tableofcontents

\section{Introduction}

Quantum spin liquids represent a class of exotic quantum phases of matter beyond the traditional Landau symmetry-breaking paradigm. Besides being conceptually interesting and experimentally relevant on their own~\cite{balents_2010,savary_2017,Zhou_QSL_review}, they  are also connected to various deep problems ranging from high-temperature superconductivity to topological order and strongly coupled gauge theories, to name a few~\cite{LeeNagaosaWen,wenbook}.

A particularly interesting quantum spin liquid in two spatial dimensions is the Dirac spin liquid (DSL)~\cite{LeeNagaosaWen,hastings_2000,hermele_2005_mother,ran_2007,hermele_2008,he_2017,Iqbal_triangular,zhu_2018}. The DSL is described by fermionic spinons -- emergent particles carrying spin-$1/2$ -- whose dispersion at low energies is described by the massless Dirac equation. These Dirac spinons interact with an emergent photon ($U(1)$ gauge field), an effective field theory known as QED$_3$. This spin liquid state was originally discussed on the square lattice in the context of high-Tc cuprates\cite{LeeNagaosaWen} and as a ``mother state" of different competing orders\cite{hermele2005}. On the kagom\'e lattice the DSL is a candidate ground state for the  Heisenberg antiferromagnet~\cite{ran_2007,hermele_2008,Iqbal_kagome}, as supported by recent DMRG calculations~\cite{he_2017,zhu_2018}, and may potentially be relevant for experimental systems such as hertbertsmithite \cite{Kagomeexpt,Norman}, although gapped spin liquids have also been proposed in this context.
 On the triangular lattice, a spin liquid is observed in  DMRG studies when a small second neighbor spin coupling $J_2$ is added in the range $0.07<J_2/J_1<0.15$~\cite{Zhu_triangular}, and variational Monte Carlo simulation suggested it to be a Dirac spin liquid~\cite{Iqbal_triangular}. As predicted for a Dirac spin liquid, a chiral spin liquid is obtained in this parameter range as soon as a time reversal symmetry breaking perturbation is applied~\cite{Sheng,Lauchli}. 
Further support comes from recent lattice gauge theory simulations have reported that QED$_3$, even with relatively small number of fermion flavors, may exhibit a stable critical phase, at least when symmetry lowering perturbations and monopoles were absent (sometimes called `non-compact'  QED$_3$)~\cite{qedcft0,qedcft}. This raises the remarkable possibility that the DSL may be realized as a stable phase (or perhaps a critical point, see ~\cite{jian_2017}) on the triangular lattice. Intriguingly, quantum spin liquid materials candidates have recently emerged on the triangular lattice \cite{triangle_sl, Law6996}. 

To make progress however, one really needs a rigorous understanding of  monopoles~\cite{polyakov_1977,hermele_20041,kapustin_2002}, an  important class of excitations (or more accurately critical fluctuations), in the DSL. 
If symmetry-allowed and relevant (in the renormalization-group sense), the proliferation of monopole instantons leads to instabilities of the DSL~\cite{shortpaper}. The properties of the monopoles also decide the nature of other, more conventional, phases in proximity to the DSL~\cite{shortpaper}. However, a fundamental aspect of the monopoles, their quantum numbers under the microscopic symmetries (lattice, time-reversal, etc.), has long been an unresolved issue. In the simpler case of a semiclassical theory of fluctuating Neel order, or equivalently a theory based on {\em bosonic} spinons (Schwinger bosons) coupled to a U(1) gauge field that naturally appears on bipartite lattices, the lattice symmetry properties of monopoles  were calculated in~\cite{HaldaneBerry,ReSaSUN}. This played an important role in predicting valence bond solid (VBS) order as a competing singlet state, and in the development of deconfined criticality~\cite{senthil_20031,senthil_20041}, of the Neel-VBS phase transiton. However, for fermionic spinons which provides an intrinsically quantum mechanical description, such an analytic understanding is still absent. Some progress was made in Ref.~\cite{alicea_2008}, in which the monopole quantum numbers on square lattice were shown to be constrained by group-theoretic considerations~\cite{AliceaFVortex} and were eventually calculated numerically. Subsequently an analysis  of the honeycomb~\cite{ran_2008} and kagom\'e~\cite{hermele_2008} lattices were also initiated. We report the numerical computation of  monopole quantum numbers for several symmetries on these lattices in a parallel paper\cite{shortpaper}, and also discuss consequences for the stability and the phenomenology of the DSL. In contrast, in this work we uncover a close and unexpected connection between the symmetry properties of monopoles and fermion band topology.  This allows us to build on recent progress understanding band topology protected by crystalline symmetries, to develop a systematic analytical approach to calculate the monopole symmetry quantum numbers on essentially any lattice -- although we focus on the physically relevant ones including square, honeycomb, triangular and kagom\'e. 
 Armed with this deeper understanding and analytical machinery, we are able to obtain a complete understanding of the symmetry action on monopoles. Along with several new results, we clarify some misconceptions in previous work, and also verify consistency with generalized Lieb-Shultz-Mattis theorems.

Our new understanding was enabled by developments in the theories of topological insulators and topological crystalline insulators over the past decade. Essentially, the symmetry properties of the monopoles are fixed by the ``band topology" of the underlying fermionic spinons. By establishing the precise connection between band topology and monopole quantum numbers, the latter can be calculated using the technology of topological band theory. An analogous approach has long been used to determine monopole quantum numbers associated with continuous, on-site symmetries. For example, when fermions fill a Chern band, a Chern-Simons term is generated that represents charge-flux attachment. Similarly, the $S_z$-spin is carried by the monopole (a flux-quanta) in the presence of a quantized spin Hall conductance\cite{ran_2008}. In this work we leverage the full power of topological band theory to determine monopole quantum numbers to include lattice symmetries and time-reversal.

It turns out that that the monopoles' lattice momenta and angular momenta (the most challenging part of the problem) is related to an old concept in band theory: the charge (or Wannier) centers of an occupied band. The basic idea is extremely simple: if a charge sits at a point in space (the Wannier center), a monopole (magnetic flux) picks up a Berry phase when moving around it. Recent developments\cite{PoIndicators,Pofragile,Cano_2018,bouhon_2018}, especially the clarification of the notion of ``fragile topology", enabled us to calculate the charge centers even when there is an obstruction to obtaining localized Wannier states. In fact, this is a frequent occurance in the states we will discuss, nevertheless we are able to obtain the location of charge centers efficiently, which then feature sites with both negative and positive charges. We note that a similar calculation has been applied to a particular spinon mean field theory on the square lattice in Ref.~\cite{thomson_2018}, where the charges could be localized on lattice sites. 

We will also see that monopoles behave very differently on bipartite (honeycomb and square) and non-bipartite (triangular and kagom\'e) lattices: on bipartite lattices there always exists a monopole that transforms trivially under all the microscopic symmetries, making it an allowed perturbation to the theory, thereby likely destabilizing the DSL; on non-bipartite lattices this does not happen, at least in the examples we considered. The difference can be traced to the fact that on bipartite lattices one can continuously tune the DSL state to another spin liquids state with $SU(2)$ (instead of $U(1)$) gauge group. This connection leads to a different, and simpler, way of calculating monopole quantum numbers on bipartite lattices, with results that are consistent with the band topology approach.

The rest of this paper is organized as follows. In Sec.~\ref{generalities} we review for completeness, aspects of $U(1)$ Dirac spin liquids, and define precisely the problem of monopole quantum numbers. We also derive some general results on time-reversal and reflection symmetries. In Sec.~\ref{MQNBipartite} we calculate the monopole quantum numbers on bipartite lattices (honeycomb and square) by  two different methods -- with give identical results. In Sec.~\ref{wanniercenter} we develop a more general method, based on spinon charge centers, that is applicable to both bipartite and  non-bipartite lattices. In Sec.~\ref{calculation} we apply this method to calculate monopole quantum numbers for the DSL on the triangular and kagom\'e lattices. In Sec.~\ref{anomalyLSM} we discuss the connection between  some of the monopole quantum numbers and generalized  Lieb-Schultz-Mattis theorems and their corresponding field theory anomalies. We conclude in Sec.~\ref{Discussion} and discusses open issues. The spinon mean-field theory used on four types of lattices and the fermion bilinear transformations are relegated to Appendices.

\section{Generalities}
\label{generalities}

\subsection{$U(1)$ Dirac spin liquid and monopole operators}
We start with the standard parton decomposition of the spin-$1/2$ operators on the lattice
\be
\label{eqn:parton}
\vec{S}_i=\frac{1}{2}f^{\dagger}_{i,\alpha}\vec{\sigma}_{\alpha\beta}f_{i,\beta},
\ee
where $f_{i,\alpha}$ is a fermion (spinon) on site $i$ with spin $\alpha\in\{\uparrow,\downarrow\}$ and $\vec{\sigma}$ are Pauli matrices. This re-writing is exact if we stay in the physical Hilbert space, defined by the constraint $\sum_{\alpha}f_{i,\alpha}^{\dagger}f_{i,\alpha}=1$. We now relax the constraint and allow the fermionic spinons to hop on the lattice (for more details see Ref.~\cite{wenbook}), according to a mean field Hamiltonian
\be
\label{eqn:ansatz}
H_{MF}=-\sum_{ij}f^{\dagger}_it_{ij}f_j.
\ee

There is a gauge redundancy $f_i\to e^{i\alpha_i}f_i$ in the parton decomposition Eq.~\eqref{eqn:parton}, which results in the emergence of a dynamical $U(1)$ gauge field $a_{\mu}$ that couples to the fermions $f$. Each site carries a gauge charge $q_i=\sum_{\alpha}f_{i,\alpha}^{\dagger}f_{i,\alpha}-1$. In the strong coupling limit where the gauge field simply enforces a constraint $q_i=0$ on each site, the physical spin Hilbert space is recovered. However if the gauge coupling does not flow to infinity in the low energy limit (this can almost be viewed as the definition of a spin liquid phase), the gauge charge only needs to vanish on average $\langle q\rangle=0$.

We now arrange the hopping amplitudes $t_{ij}$ in a way so that the spinons form four Dirac cones at low energy: two Dirac valleys per spin, as required by fermion doubling. For example, on honeycomb lattice one can just take a uniform, real, and non-bipartite hopping, and two Dirac valleys will generically appear. The non-bipartite nature (second-neighbor hopping) is needed to make sure that the gauge group is $U(1)$ rather than $SU(2)$\cite{wenbook} -- this will play an important role later in Sec.~\ref{MQNQCD}. The relevant mean field Hamiltonians on square, honeycomb, triangle and kagom\'e lattices are described in detail in Appendix~\ref{bilinears}.

Taking the continuum limit, the $U(1)$ Dirac spin liquids in the low energy, long wavelength (IR) limit can effectively be described by the following (Euclidean) Lagrangian:
\be
\label{eqn:qedL}
\mathcal{L}=\sum_{i=1}^{4}\bar{\psi}_ii\slashed{D}_a\psi_i+\frac{1}{4e^2}f_{\mu\nu}^2,
\ee
where $\psi_i$ is a two-component Dirac fermion and $a$ is a dynamical $U(1)$ gauge field. We choose $(\gamma_0, \gamma_1, \gamma_2)=(\mu^2, \mu^3, \mu^1)$ where $\mu$ are Pauli matrices.
This theory is also known as QED$_3$ with $N_f=4$. The theory flows to strong coupling at energy scale below $e^2$, and its ultimate IR fate is not completely known. In this work we assume that when monopole instantons are suppressed (to be explained in more details below), this QED theory flows to a stable critical fixed point in the IR, as supported by recent numerics\cite{qedcft0,qedcft}.

Naively there is a conserved current in the theory
\be
j_{\mu}=\frac{1}{2\pi}\epsilon_{\mu\nu\lambda}\partial_{\nu}a_{\lambda},
\ee
that corresponds to a global $U(1)$ symmetry sometimes called $U(1)_{top}$. The conserved charge is simply the magnetic flux of the emergent $U(1)$ gauge field. One can then define operators that carries this global $U(1)_{top}$ charge, i.e. operators that create or annihilate total gauge flux by $2\pi$. We denote these operators by $\mathcal{M}$ -- one can pictorially think of it as a point in space-time surrounded by a $2\pi$-flux. This operator is not included in Eq.~\eqref{eqn:qedL}, but in principle may be included as a perturbation which explicitly breaks the $U(1)_{top}$ symmetry. In the absence of the gapless Dirac fermions (or other matter fields), it is known that such a perturbation\cite{polyakov_1977} will open a gap for the Maxwell photon and confine gauge charges. With gapless matter fields (like the Dirac fermions here) the effect of monopole perturbation is more subtle: at large enough $N_f$ (fermion flavor) the monopole becomes an irrelevant perturbation, but the lower critical $N_f$ is not completely known (some bounds were estimated from F-theorem~\cite{TarunF}).

Let us look at the monopoles in more detail. It is helpful to think in the large-$N_f$ limit, where gauge fluctuations are suppressed. In this case the monopole simply creates a static $2\pi$ magnetic flux in which the Dirac fermions move freely. The most relevant monopole corresponds to the ground state of these Dirac fermions, with all negative-energy levels filled and all positive-energy levels empty\footnote{All these can be made more precise through radial quantization and the state-operator correspondence, but for the purpose of this work we do not need to use those machineries.}. However, each Dirac cone also contributes to a zero-energy mode (guaranteed by the Atiyah-Singer index theorem) in a $2\pi$ flux background. The filling of any of these four zero modes do not affect energetics. However, gauge-invariance (i.e. vanishing of the overall gauge charge) requires that exactly half of the zero modes to be filled\cite{kapustin_2002}. This gives in total $C_4^2=6$ distinct (but equally relevant) monopoles, schematically written as
\be
\label{eqn:Mzero}
\Phi\sim f^{\dagger}_if^{\dagger}_j\mathcal{M}_{bare},
\ee
where $f^{\dagger}_i$ creates a fermion in the zero-mode associated with $\psi_i$, and $\mathcal{M}_{bare}$ creates a ``bare" flux quanta without filling any zero mode.
For later convenience, we define the six monopoles as
\begin{align}
\label{eqn:monopole_type}
\Phi^\dagger_{1/2/3}=f^\dagger_{\alpha,s} (\epsilon \tau^{1/2/3})^{\alpha\beta} \epsilon^{ss'} f^\dagger_{\beta,s'}\mathcal{M}_{bare}\nonumber\\
\Phi^\dagger_{4/5/6}=i f^\dagger_{\alpha,s} (\epsilon )^{\alpha\beta}( \epsilon\sigma^{1/2/3})^{ss'} f^\dagger_{\beta,s'}\mathcal{M}_{bare}
\end{align}
where we refine the label of zero mode by valley indices $\alpha=1,2$ and spin indices $s=\up,\down$, $\epsilon$ is the antisymmetric rank-$2$ tensor, which is necessary because of the anticommutation relations of $f$ operators, $\tau,\sigma$ acts on valley/spin indices as the standard Pauli matrices formalism. The factor $i$ in the second line is necessary such that the six monopoles are related by $SU(4)$ rotations of Dirac fermions (to be explained in more detail later). From our construction, it's straightforward to see that the first three monopoles are spin singlets, while the latter three monopoles are spin triplets.
Both Dirac sea and the zero modes contribute to properties of monopoles such as symmetry quantum numbers.

One can likewise define ``anti-monopoles" as operators inserting $-2\pi$ flux and appropriately filling Dirac zero modes. However, it is more convenient for us to simply view such operators as the ``anti-particles", or hermitian conjugates, of the monopole operators defined above.

Notice that under a $2\pi$-flux, each zero-mode behaves as a Lorentz scalar \cite{kapustin_2002}(since there is no other index responsible for higher spin), in contrast to its parent Dirac fermion (originally a spinor). This makes the monopole operator also a Lorentz scalar.

\subsection{Symmetries}

We now carefully examine the global symmetries of the continuum QED$_3$ theory.  Clearly we have the Lorentz symmetry, together with the standard charge conjugation $\mathcal{C}$, time-reversal $\mathcal{T}$ and space reflection $\mathcal{R}_x$. As we have discussed already, the conservation of the gauge flux corresponds to a global $U(1)$ symmetry known as the topological $U(1)_{top}$. The fermion flavor symmetry is naively $SU(4)$: $\psi_i\to U_{ij}\psi_j$ where $U\in SU(4)$, but we should remember that global symmetries, properly defined, should only act on gauge invariant local operators. Naively one would consider fermion bilinear operators like $\bar{\psi}\sigma^{\mu}\tau^{\nu}\psi$ as the simplest gauge-invariant operators, which transform as ($15$-dimensional) $SU(4)$ adjoints. However, the monopole operators (defined in Eq.~\eqref{eqn:Mzero} or \eqref{eqn:monopole_type}) transform as a $6$-dimensional vector under $SU(4)$, or more precisely $SO(6)=SU(4)/\mathbb{Z}_2$. Notice that this operator is odd under both the $SO(6)$ center ($-I_{6\times6}$) and a $\pi$-rotation in $U(1)_{top}$. More generally one can show that any local operator has to be simultaneously odd or even under the two operations -- for example a fermion bilinear $\bar{\psi}_i\psi_j$ carries no gauge flux and is even under the $SO(6)$ center. So the proper global symmetry group should be
\be
\frac{SO(6)\times U(1)_{top}}{\mathbb{Z}_2}
\ee
together with $\mathcal{C, T, R}_x$ and Lorentz. One can certainly consider $2\pi$-monopoles in higher representations of $SO(6)$, but in this work we will assume that the leading monopoles (with lowest scaling dimension) are the ones forming an $SO(6)$ vector -- this is physically reasonable and can be justified in large-$N_f$ limit.

Instead of working with the explicit definition of monopoles from Eq.~\eqref{eqn:Mzero}, we shall simply think of the monopoles as six operators $\{\Phi_1,...\Phi_6\}$ that carry unit charge under $U(1)_{top}$ and transform as a vector under $SO(6)$: $\Phi_i\to O_{ij}\Phi_j$. Likewise we define ``anti-monopoles" as six operators $\{\Phi_1^{\dagger},...\Phi_6^{\dagger}\}$ that also transform as an $SO(6)$ vector, but carry $-1$ charge under $U(1)_{top}$. The virtue of defining the monopole operators abstractly based on symmetry representations is that we can easily fix the $\mathcal{C, T, R}_x$ symmetry actions on the monopoles completely based on the group structure. Consider a ``bare" time-reversal symmetry
\be
\label{eqn:T0}
\mathcal{T}_0: \psi\to i\gamma_0\sigma^2\tau^2\psi,
\ee
where $\gamma$ acts on the Dirac index, $\sigma$ acts on the physical spin index, and $\tau$ acts on the ``valley" index. The physical time-reversal symmetry $\mathcal{T}$ (to be discussed later) is in general a combination of $\mathcal{T}_0$ and some additional $SU(4)$ rotation $U_T$. Now consider the group structure of the $SO(6)\times U(1)_{top}/\mathbb{Z}_2$ symmetry and $\mathcal{T}_0$. Clearly $\mathcal{T}_0$ commutes with $U(1)_{top}$, which simply means that monopoles become anti-monopoles. $\mathcal{T}_0$ also commutes with $SU(4)$ rotations generated by $\{\sigma^{i},\tau^{i}\}$, namely the spin-valley subgroup $SO(3)_{spin}\times SO(3)_{valley}$. But for those generated by $\{\sigma^{i}\ \tau^j\}$ we have $\mathcal{T}_0U=U^{\dagger}\mathcal{T}_0$. One can then show that the only consistent implementations on the monopoles are $\mathcal{T}_0: \Phi\to \pm O_T\Phi^{\dagger}$, where
\be
\label{eqn:OT}
O_T=\left(\begin{array}{cc} I_{3\times 3} & 0 \\ 0 & -I_{3\times 3}   \end{array}\right).
\ee
The basis is chosen so that $\Phi_{1,2,3}$ rotates under the $SO(3)$ generated by $\tau^{i}$, and $\Phi_{4,5,6}$ rotates under that by $\sigma^{i}$. Importantly, $O_T\in O(6)$ but not $SO(6)$. One can likewise consider a ``bare" reflection
\be
\label{eqn:R0}
\mathcal{R}_{0}: \psi(x)\to i\gamma_1\psi(\mathcal{R}x).
\ee
Since this symmetry commutes with $SO(6)$ rotations but flips $U(1)_{top}$ charge, we have for the monopoles $\mathcal{R}_{0}: \Phi_i\to \Phi_i^{\dagger}$ (up to a phase factor which can be absorbed through a re-definition of $\Phi$). Finally for the ``bare" charge-conjugation
\be
\label{eqn:C0}
\mathcal{C}_0: \psi\to \sigma^2\tau^2\psi^{*}.
\ee
We notice that it has the same commutation relation to $SO(6)$ as $\mathcal{T}_0$ and also flips $U(1)_{top}$ charge. Therefore $\mathcal{C}_0: \Phi\to \pm O_T\Phi^{\dagger}$. {This analysis also shows that the fermion mass operators $\bar{\psi}_iT_{ij}\psi_j$ that form an adjoint representation of $SU(4)$ ($T$ is an $SU(4)$ generator) is indistinguishable from $i\Phi_i^{\dagger}A_{ij}\Phi_j$ in terms of symmetry quantum numbers, where $A$ is a real $6\times 6$ anti-symmetric matrix.}

Clearly the lattice spin Hamiltonians would not have the full continuum symmetry -- typically we only have spin rotation, lattice translation and rotation, reflection and time-reversal symmetries. It was argued in Ref.~\cite{hermele2005} that the enlarged symmetry (such as $SO(6)\times U(1)_{top}/\mathbb{Z}_2$) would emerge in the IR theory since terms breaking this symmetry down to the microscopic ones are likely to be irrelevant (justified in large-$N_f$ analysis). In this work we will assume that the enlarged symmetry does emerge, at least before the monopole tunnelings are explicitly added to the Lagrangian.

The central question in this paper is: given a realization of a $U(1)$ Dirac spin liquid on some lattice, how do the monopoles transform under the microscopic symmetries (such as lattice translation)? Since we already know how the monopoles transform under the IR emergent symmetries (such as $SO(6)\times U(1)/\mathbb{Z}_2$), the question can be equivalently formulated as: how are the microscopic symmetries embedded into the enlarged symmetry group? Clearly spin-rotation can only be embedded as an $SO(3)$ subgroup of the $SO(6)$ flavor group, meaning that three of the six monopoles ($\Phi_{4,5,6}$ from Eq.~\eqref{eqn:monopole_type}) form a spin-$1$ vector, and the other three ($\Phi_{1,2,3}$ from Eq.~\eqref{eqn:monopole_type}) are spin singlets. Other discrete symmetries can be realized, in general, as combinations of certain $SO(6)$ rotations followed by a nontrivial $U(1)_{top}$ rotation, and possibly some combinations of $\mathcal{C}_0, \mathcal{T}_0, \mathcal{R}_0$. (Remember that Lorentz group acts trivially on the $2\pi$-monopoles.) In fact, in all the examples we are interested in, all those discrete symmetries commute with the spin $SO(3)$ rotation. This means that for most purposes we can focus on the $SO(3)_{spin}\times SO(3)_{valley}$ subgroup of $SO(6)$, and the realization of the discrete symmetries should only envolve $SO(3)_{valley}$ and possibly $\mathcal{C,P,T}$.

Many of these group elements in a symmetry realization can be fixed from the symmetry transformations of the Dirac fermions $\psi_i$, which is in turn fixed by the symmetry of the mean field ansatz in Eq.~\eqref{eqn:ansatz}, under the name of projective symmetry group (PSG)\cite{wenbook}. For example, if the symmetry operation acts on $\psi$ as $\psi\to U\psi$ with a nontrivial $U\in SU(4)$, then we know that the monopoles should also be multiplied by an $SO(6)$ matrix $O$ that corresponds to $U$. This $SO(6)$ matrix $O$ can be uniquely identified up to an overall sign, which can also be viewed as a $\pi$-rotation in $U(1)_{top}$. In practice the $SO(6)$ flavor rotation involved in a symmetry realization is always within the $SO(3)_{spin}\times SO(3)_{valley}$ subgroup. The six monopoles transform as $(1,0)\oplus (0,1)$ under this subgroup, which is the same representation of the six fermion bilinears $\{\bar{\psi}\sigma^i\psi,\bar{\psi}\tau^i\psi\}$. Therefore to fix the $SO(6)$ rotation of the monopoles in a given symmetry realization it is sufficient to fix that for the six masses, also known as quantum spin Hall and quantum valley Hall masses, respectively. For the examples we are interested in, these information are also reviewed in Appendix.~\ref{bilinears}.

Similar logic applies to operations like $\mathcal{C, T, R}_x$. The only exception is the flux symmetry $U(1)_{top}$: there is no information regarding $U(1)_{top}$ in the PSG. Fixing the possible $U(1)_{top}$ rotations in the implementations of the microscopic discrete symmetries is the main task of this work.

The difficulty in fixing the $U(1)_{top}$ factor in a symmetry transform lies in its UV nature: intuitively, the $U(1)_{top}$ phase factor comes from the Berry phase accumulated when a monopole moves on the lattice scale, in a nontrivial ``charge background"\cite{HaldaneBerry, ReSaSUN}. This lattice-scale feature is not manifested directly in simple objects in the IR (such as the Dirac fermions). In previous studies such phase factors were decided numerically\cite{alicea_2008,ranvishwanathlee,shortpaper}. In this paper we will develop several different analytical methods to calculate such phase factors.

We should emphasize here that the questions addressed in this work are kinematic (rather than dynamical) in nature: i.e. we are  interested in the qualitative properties of the monopoles, such as symmetry representations, rather than quantitative properties such as scaling dimensions. We will introduce, at various stages of our argument, assumptions that are only justified in certain limits (such as at  large-$N_f$), and importantly although these assumptions will not be completely satisfied, they will provide a rationale for selecting an answer typically from a discrete set of possibilities. Some of the particularly important assumptions that follow from this treatment are: (1) the most relevant monopole operator are those that transform as an $SO(6)$ vector and Lorentz scalar -- this is physically reasonable but justifiable only in large-$N_f$, (2) when perturbed by an adjoint mass $\bar{\psi}\sigma^{\mu}\tau^{\nu}\psi$, the small mass limit is adiabatically connected with the large mass limit which describes lattice scale physics and the Wannier limit (roughly speaking, this means that the adjoint mass is not only relevant, but flows all the way to infinity in the IR)\footnote{ The flow to infinity is smooth with no singularity/fixed point.} -- another physically reasonable assumption that is justified in large-$N_f$, and (3) the $U(1)_{top}$ phase factors in the microscopic symmetry realizations are decided completely by the mean-field theory of the spinons (which is a free fermion theory) -- gauge fluctuations only modify other quantitative features (such as scaling dimensions) but not the (discrete) symmetry properties. In particular, assumption (3) may not always be valid (depending on microscopic details), but when it is not valid, the  parton construction combined with the  mean-field description is itself not likely  to provide a reasonable starting point to describe the phase.

\subsection{Time-reversal, reflection and band topology}
\label{dimred}

We now derive some general results for time-reversal and reflection symmetries that will be generally applicable in all the systems we are interested in. Essentially, with the help of the exact $SO(3)$ spin rotation symmetry, the $U(1)_{top}$ phase factors associated with time-reversal and reflection can be uniquely fixed.

Since the $U(1)_{top}$ phase factor comes from UV physics, we can deform the QED$_3$ theory with a fermion mass to make the theory IR trivial, so that we can focus on the UV part. Consider perturbing with a mass term
\be
\label{eqn:adjointmass}
\Delta\mathcal{L}=m\bar{\psi}_iT_{ij}\psi_j,
\ee
where $T_{ij}$ is chosen to be an $SU(4)$ generator without loss of generality. The fermions are now gapped, and there is a pure Maxwell $U(1)$ gauge theory left in the IR. The zero-modes associated with the monopoles are lifted (according to the signs of the fermion masses), lifting the six-fold degeneracy of the monopole completely, so that there is only one gapless monopole left in the Maxwell theory. The identity of the surviving monopole is fixed again by symmetry. The mass term breaks the flavor symmetry from $SO(6)$ to $SO(4)\times SO(2)$. If we probe the theory with an $SO(4)\times SO(2)$ gauge field $\mathcal{A}^{SO(4)}+A^{SO(2)}$, the massive fermions generate a topological term
\be
\label{eqn:qshtop}
\mathcal{L}_{top}=\frac{{\rm sgn}(m)}{2\pi}A^{SO(2)}da.
\ee
This means that the monopole (now unique) carries $\pm 1$ charge under the $SO(2)$ generated by $T$ in Eq.~\eqref{eqn:adjointmass} and is a singlet under the remaining $SO(4)$ -- this uniquely fixes the identity of the monopole among the six degenerate ones in the gapless phase. The argument will also be useful for deciding the nature of the symmetry-breaking phase when a mass perturbation is turned on, since the monopole will eventually spontaneously condense in the Maxwell theory, possibly breaking further symmetries (such as the $SO(2)$ here).

Now as long as the relevant symmetry, let us call it $g$, is not broken by the mass perturbation, the $U(1)_{top}$ phase factor $U_{top}^g$ associated with the implementation of $g$ will not be affected. We are thus left with the simpler problem of finding the Berry phase of a non-degenerate monopole moving in an insulating (gapped) charge background, which is essentially determined by the topology of the insulator. Since the other monopoles are related to this monopole by some $SO(6)$ flavor rotations, their symmetry transformations are fixed once we obtain the transformation of this particular monopole.

It turns out to be particularly useful, for all the examples to be considered in this work, to consider a ``quantum spin Hall" (QSH) mass perturbation
\be
\Delta\mathcal{L}_{QSH}=m\bar{\psi}\sigma^3\psi,
\ee
where $\sigma^3$ acts in the spin but not valley index. This term breaks the spin $SO(3)$ rotation down to $SO(2)$ and generates a mutual spin-charge Hall response as in Eq.~\eqref{eqn:qshtop}, so that the low energy monopole operator transforms as $S_z={\rm sgn}(m)$ -- in our notation this monopole is denoted as $\Phi_4\pm i\Phi_5$. Crucially, this term leaves all other discrete symmetries unbroken -- except for reflection symmetries $\mathcal{R}$, but it is still a symmetry when combined with a spin-flip operation $\RR'=\sigma^2\RR$. Since the $U(1)_{top}$ factors can be nontrivial only for lattice  symmetries and time reversal, they can all be determined by considering monopole quantum numbers in this QSH insulator. For example, it is well known that the QSH insulator is also a $\mathbb{Z}_2$ topological insulator\cite{KaneMele}, protected by the Kramers time-reversal symmetry $\mathcal{T}: f\to i\sigma^2f$. This fixes the transformation of the monopole under time-reversal to be:
\be
\label{eqn:monopoleTR}
\mathcal{T}: \mathcal{M}\to -\mathcal{M}^{\dagger},
\ee
where the non-triviality of the topological insulator is manifested in the minus sign. This can be seen most easily through the edge state of the QHE insulator
\be
\label{eqn:edge}
H_{edge}=\int dx \chi^{\dagger}i\sigma^2\partial_x\chi,
\ee
where $\chi$ is a two-component Dirac fermion in $(1+1)d$, with time-reversal acting as $\chi\to i\sigma^2\chi$ that forbids any mass term. In the QHE state a monopole tunneling event will transfer one left-moving fermion into a right-moving one, which is nothing but the physics of axial anomaly. This leads to the operator identification on the edge $\mathcal{M}\sim \chi^{\dagger}(\sigma^1+i\sigma^3)\chi$, from which Eq.~\eqref{eqn:monopoleTR} follows.

Reflection symmetry can be discussed in a similar manner. If the reflection does not involve charge conjugation, the monopole simply transforms as
\be
\mathcal{R}_x: \mathcal{M}\to e^{i\theta_{\mathcal{R}}}\mathcal{M}^{\dagger}.
\ee
where the overall phase factor can change by a re-definition of $\mathcal{M}$ and is therefore physically meaningless -- unless there are more than one reflection axis, in which case the relative phases in the transforms become meaningful. Charge conjugation symmetry (if exists) alone is not interesting for the same reason.

It becomes more interesting, and turns out to be also much simpler, when a reflection involves an extra charge conjugation operation, denoted as $\CC\RR$. Under this symmetry, the monopole is mapped to itself, possibly with a sign
\be
\label{eqn:CRmonopole}
\mathcal{CR}: \mathcal{M}\to \pm \mathcal{M},
\ee
where we have assumed that $(\mathcal{CR})^2=1$ on local operators. The two different signs in the above transformation are physically distinct and represent different topology of the underlying insulators. 

In a quantum spin Hall insulator, it turns out to be particularly simple to tell if the insulator is also nontrivial under an additional $\mathcal{CR}$: it is trivial if $(\mathcal{CR})^2\psi=-\psi$, and nontrivial if $(\mathcal{CR})^2\psi=+\psi$ (which then leads to the nontrivial sign in Eq.~\eqref{eqn:CRmonopole}). In particular, the ``bare" $\mathcal{CR}$ symmetry defined in Eq.~\eqref{eqn:C0} and \eqref{eqn:R0} squares to one and therefore has nontrivial transformation.

The easiest way to understand the above statement is to consider the edge state Eq.~\eqref{eqn:edge} that preserves both charge, spin-$S_z$ and $\mathcal{CR}$ symmetry. There are two different ways to implement a $\mathcal{CR}$ symmetry: $\mathcal{CR}_+: \chi(x)\to \chi^{\dagger}(-x)$ or $\mathcal{CR}_-: \chi(x)\to \sigma^2\chi^{\dagger}(-x)$, where $(\mathcal{CR}_+)^2=1$ and $(\mathcal{CR}_-)^2=-1$ on the fermions. It is easy to see that $\mathcal{CR}_+$ forbids a Dirac mass term, making the insulator also nontrivial under $\mathcal{CR}_+$, while $\mathcal{CR}_-$ does not forbid any mass term and is therefore trivial. The monopole transformation under $\mathcal{CR}$ in Eq.~\eqref{eqn:CRmonopole} can be obtained through the operator identification $\mathcal{M}\sim \chi^{\dagger}(\sigma^1+i\sigma^3)\chi$. This result is perhaps natural when we think of $\mathcal{CR}$ as obtained from Wick-rotating time-reversal symmetry $\mathcal{T}$ using $\CC\RR\T$ theorem\cite{wittenreview}, where $\mathcal{T}^2=(-1)^F$ rotates to $(\CC\RR)^2=+1$.  

It is now natural to ask what would happen if we had a ``quantum valley Hall" mass $\bar{\psi}\tau^i\psi$ instead. Now since the full spin $SO(3)$ is unbroken, the insulator cannot have the band topology of topological insulators (under either $\mathcal{T}$ or $\mathcal{CR}$) -- this is simply the famous statement that topological insulator requires spin-orbit coupling. Therefore time-reversal and reflection (or $\mathcal{CR}$) should act trivially on spin singlet monopoles selected by the valley Hall masses.\footnote{One might wonder why our previous argument for the quantum spin Hall insulator cannot be used for the quantum valley Hall insulator -- afterall the time-reversal action $\mathcal{T}_0$ appeared to be democratic between valley and spin indices. Crucially, in our systems the continuous valley symmetry is never exact in the UV, but can only be emergent in the IR. This has to do with a mixed anomaly between the valley $SO(3)_{valley}$ and spin $SO(3)_{spin}$ in the continuum field theory, which is one manifestation of the parity anomaly. This means that one cannot use the previous argument here, since the link between quantum spin Hall and time-reversal topological insulating behavior relies crucially on the exactness of the spin rotation symmetry -- in particular, the edge state argument assumes that there is no anomaly associated to the symmetries involved (a point dubbed ``edgability" in Ref.~\cite{WS2013}). 
}

To summarize, the two symmetries $\mathcal{T}_0$ and $\mathcal{CR}_0$ defined in Eq.~\eqref{eqn:T0},~\eqref{eqn:C0} and \eqref{eqn:R0} acts on the six monopoles as
\bea
\label{T0CR0}
\mathcal{T}_0: && \left(\begin{array}{c}
    \Phi_{1,2,3}\\
    \Phi_{4,5,6}
\end{array} \right) \to \left(\begin{array}{c}
    \Phi_{1,2,3}^{\dagger}\\
    -\Phi_{4,5,6}^{\dagger}
\end{array} \right), \nn 
\mathcal{CR}_0: && \left(\begin{array}{c}
    \Phi_{1,2,3}\\
    \Phi_{4,5,6}
\end{array} \right) \to \left(\begin{array}{c}
    \Phi_{1,2,3}\\
    -\Phi_{4,5,6}
\end{array} \right).
\eea
The physical time-reversal and reflection may further involve additional $SO(6)$ rotations or charge conjugations, which can be included straightforwardly.

Next we turn to the more complicated symmetries including lattice translation and rotations. We first discuss the simpler cases on bipartite lattices.

\section{Monopole quantum numbers I: bipartite lattices}
\label{MQNBipartite}
\subsection{Monopole quantum numbers constrained by QCD$_3$}
\label{MQNQCD}

On bipartite lattices, at least for the examples considered in this work, we can always continuously tune the  mean field Hamiltonian Eq.~\eqref{eqn:ansatz}, without breaking any symmetry or changing the low-energy Dirac dispersion, to reach a point with particle-hole symmetry:
\be
\CC: f_{i,\alpha}\to (-1)^ii\sigma^2_{\alpha\beta}f^{\dagger}_{i,\beta}.
\ee
This theory will then have a larger gauge symmetry of $SU(2)_g$, with $(f_{\alpha},i\sigma^y_{\alpha\beta}f^{\dagger}_{\beta})^T$ forming an $SU(2)_g$ fundamental (anti-fundamental) on each site in A-sublattice (B-sublattice) for each spin $\alpha$. The low energy theory again has four Dirac cones, with two valleys, each forming a fundamental under both gauge $SU(2)_g$ and spin $SU(2)_s$. The continuum field theory of such state, described by an $SU(2)$ gauge field coupled to four Dirac cones, is also known as QCD$_3$ with $N_f=2$. The Lagrangian is given by
\be
\mathcal{L}_{QCD}=\sum_{i=1}^{2}\bar{\chi}_i(i\slashed{\partial}+\slashed{a}^{SU(2)})\chi_i.
\ee

At the lattice scale the $SU(2)$ gauge symmetry can be Higgsed down to $U(1)$ by reinstating the  particle-hole symmetry breaking hopping (which could be weak), and our familiar $U(1)$ Dirac spin liquid will be recovered at low energy. However, it turns out to be very useful to consider an intermediate theory in which the $SU(2)$ gauge symmetry is Higgsed down to $U(1)$, but the particle-hole symmetry survives as a global $\mathbb{Z}^C_2$ symmetry. At low energy this theory also flows to QED$_3$, but with an extra $\mathbb{Z}_2^C$ symmetry compared to the $U(1)$ Dirac spin liquid. In fact this theory does not faithfully represent a lattice spin system due to the extra $\mathbb{Z}^C_2$. However, it is a perfectly well-defined lattice gauge theory, and can represent a lattice spin system if the spin-rotation symmetry is enlarged from $SO(3)$ to $O(3)\sim SO(3)\times\mathbb{Z}^C_2$.  In the continuum theory this intermediate QED$_3$ can be obtained from the QCD$_3$ theory by condensing a Higgs field that carries spin-$1$ of the gauge $SU(2)$ and is  even under $\mathbb{Z}_2^C$. This Higgs condensation does not break any global symmetry of the QCD$_3$ and the low energy Dirac dispersion is not affected. 
We can then safely view the continuum QCD$_3$ field theory (which is free in the UV), instead of the original lattice theory, as the UV completion of the intermediate QED$_3$ theory. 
As we shall see below, the virtue of this alternative UV completion is that the QCD$_3$ theory is much easier to understand than the full lattice theory.

The QCD theory has the standard Lorentz and $\mathcal{T, R}_x$ symmetries. It may not necessarily flow to a conformal fixed point\cite{KNQCD}, but this should not be important for our discussion since we are not interested in the ultimate IR fate of this theory. The flavor symmetry of QCD$_3$ at $N_f=2$ is $SO(5)$, which acts on the fermions as $Spin(5)=Sp(4)$. Crucially, there is no additional topological symmetry since the flux of $SU(2)$ gauge field is not conserved. 

Now we notice that the implementation of the microscopic (continuous or discrete) symmetries in the QCD$_3$ theory is completely fixed by the symmetry transform of the Dirac fermions $\chi$, due to the absence of any gauge flux conservation. For example, a nontrivial $Sp(4)$ transform on $\chi$ maps to a unique $SO(5)$ transform on gauge-invariant operators such as fermion bilinears. Notice that in the $SU(4)\to SO(6)$ mapping discussed in the QED$_3$ context we still had a sign ambiguity due to the existence of $SO(6)$ center $-I_{6\times6}$. That ambiguity is absent here since $SO(5)$ has no center. The bottom line is that we know completely how the microscopic symmetries are embedded into the symmetries of the continuum QCD$_3$ field theory. Once we reach the QCD description, the exact nature of these symmetries at the lattice scale is no longer important -- we simply view them as part of the $SO(5)\times Lorentz\times \mathcal{T,R}_x$ symmetry. Now we Higgs the $SU(2)$ gauge symmetry down to $U(1)$, and far below the Higgs scale we obtain the intermediate QED$_3$ theory which has a larger emergent symmetry including the $SO(6)\times U(1)/\mathbb{Z}_2$ and $\mathbb{Z}_2^C$ symmetries. 

We now show that there is a unique way to embed the symmetries of QCD ($SO(5)$ and $\mathcal{R, T}$) into the symmetries of QED. This will in turn fix the embedding of the microscopic symmetries into the symmetries of QED. First, it is obvious that there is a unique way to embed the continuous $SO(5)$ symmetry of QCD to $SO(6)\times U(1)/\mathbb{Z}_2$ of QED, up to re-ordering of operators: five of the six monopoles should transform as an $SO(5)$ vector and the remaining one (call it $\Phi_{trivial}$) should be an $SO(5)$ singlet, or $6=1\oplus 5$. Since the microscopic spin rotation symmetry $SO(3)_{spin}$ must be part of the $SO(5)$, the $SO(5)$ singlet monopole $\Phi_{trivial}$ must also be a spin singlet, i.e. it is a combination of $\Phi_{1,2,3}$. Crucially, there is no nontrivial $U(1)_{top}$ phase factor involved in the realization of the $SO(5)$ symmetry on the monopoles.

Now consider the bare time-reversal and reflection-conjugation defined in Eq.~\eqref{eqn:T0},~\eqref{eqn:C0} and \eqref{eqn:R0}. As we argued before they act on the monopoles as Eq.~\eqref{T0CR0}. This immediately implies that the $SO(5)$ singlet monopole $\Phi_{trivial}$ also transforms trivially under $\mathcal{T}_0$ and $\mathcal{CR}_0$. The physical time-reversal and reflection symmetry may involve a further $SO(5)$ flavor rotation, but this will not affect $\Phi_{trivial}$ since it is an $SO(5)$ singlet. For the charge-conjugation symmetry $\mathcal{C}$, we expect $\Phi_{trivial}\to e^{i\theta}\Phi_{trivial}^\dagger$ for some phase factor $e^{i\theta}$ since $\mathcal{C}$ cannot mix $SO(5)$ singlet with $SO(5)$ vector, and the phase can be chosen to be trivial by redefining the monopole operators. In this case ${\rm Re}\Phi_{trivial}\equiv(\Phi_{trivial}+\Phi_{tivial}^{\dagger})/2$ is trivial under $\mathcal{C}$ (in later examples sometimes the opposite convention is chosen). We then conclude that ${\rm Re}\Phi_{trivial}$ is trivial under all microscopic symmetries in the intermediate QED$_3$ theory.

We now consider the actual $U(1)$ Dirac spin liquid of interest to us. This can be obtained from the intermediate QED$_3$ by explicitly breaking the $\mathcal{C}$ symmetry. It is also possible, as we shall see on square lattice, that some other symmetries such as translation $T_{1,2}$ and time-reversal $\mathcal{T}$ are also broken, but the combinations $T_{1,2}\mathcal{C}$ and $\mathcal{T}\mathcal{C}$ are preserved. In any case, the symmetry-breaking term does not change the low-energy Dirac dispersion  (except velocity anisotropy) and is expected to be irrelevant. Therefore we do not expect any change in monopole symmetry quantum numbers -- as long as the symmetries are still unbroken. In particular, the trivial monopole $\Phi_{trivial}$, or at least ${\rm Re}\Phi_{trivial}\equiv(\Phi_{trivial}+\Phi_{tivial}^{\dagger})/2$, should still transform trivially under all global symmetries.

\begin{figure}
\captionsetup{justification=raggedright}
 \begin{center}
\adjustbox{trim={0.12\width} {0\height} {0\width} {0\height},clip}
{\includegraphics[width=1.3\columnwidth]{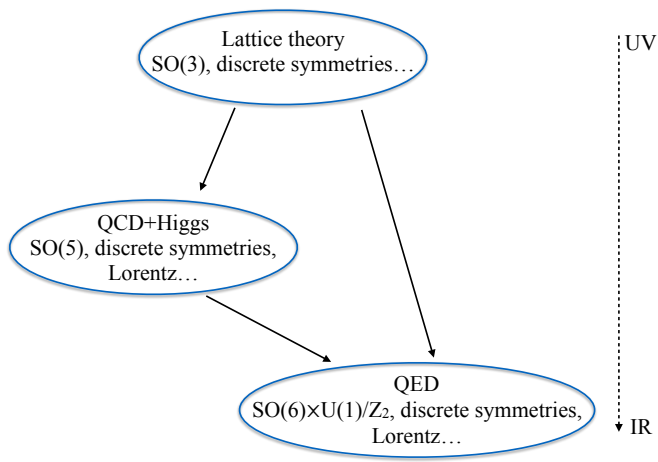}}

\end{center}
\caption{For the Dirac spin liquid on  bipartite lattices (honeycomb and square) we can view the continuum QCD$_3$ $+$ Higgs field theory (instead of the original lattice theory) as the UV completion of the QED$_3$ theory. The implementation of microscopic symmetries (at the lattice scale) in the QCD theory can be fixed by PSG analysis, while the implementation of QCD symmetries in the QED theory is also uniquely fixed by field theory analysis. This uniquely fixes the implementation of microscopic symmetries in the QED ($U(1)$ Dirac spin liquid) theory.}

\label{fig:Flow}
\end{figure}

In summary, on bipartite lattices there is always one monopole operator (at least the real or imaginary part of it) that transforms trivially under all microscopic symmetries. The reasoning is summarized in Fig.~\ref{fig:Flow}. 

Although we have emphasized the bipartieness of the lattices in our argument, the discussion above showed that what really mattered was whether the $U(1)$ spin liquid mean-field ansatz could be upgraded to an $SU(2)$ gauge theory. One could certainly consider mean-field ansatz on bipartite lattices that are ``intrinsically non-bipartite", meaning they cannot be adiabatically tuned to have an $SU(2)$ gauge structure. One could also consider ansatz on non-bipartite lattices that are compatible with $SU(2)$ gauge structure, for example by making all the hoppings imaginary (giving up time-reversal symmetry; although this would typically induce a gap). Our dichotomy on bipartiteness should be applied with care in those scenarios.

The above argument also completely fixes all the monopole quantum numbers on bipartite lattices. We now look at honeycomb and square lattices in detail.

\subsubsection{Honeycomb lattice}

The uniform hopping mean-field ansatz on honeycomb gives the $N_f=4$ QED$_3$ low-energy theory. The Dirac points stay at momenta $\mathbf K=(\frac{2\pi}{3},\frac{2\pi}{3}),\mathbf K'=-\mathbf K$.  Under appropriate basis, the low-energy effective Hamiltonian takes the standard Dirac form. The physical symmetries act as
\begin{align}
\label{eqn:honey_psg}
T_{1/2}&: \psi\rightarrow e^{-i\frac{2\pi}{3}\tau^3} \psi\quad
C_6: \psi\rightarrow -i e^{-i\frac{\pi}{6}\mu^3} \tau^1 e^{-i\frac{2\pi}{3}\tau^3} \psi \nonumber\\
R_x&: \psi\rightarrow -\mu^2\tau^2 \psi\quad
R_y: \psi\rightarrow \mu^1\tau^3 \psi \nonumber\\
\mathcal T&: \psi\rightarrow -i\sigma^2\mu^2\tau^2 \psi \quad
\mathcal C: \psi\rightarrow i\mu^1\tau^1\sigma^2\psi^*
\end{align}
where $\mu^i$ are Pauli matrices acting on the Dirac spinor index, $T_{1/2}$ is the translation along two basis vectors with $2\pi/3$ angle between them,$C_6$ is $\pi/3$ rotation around a center of a honeycomb plaquette, and $R_{x/y}$ denotes reflection perpendicular to the direction of the unit cell/ the axis perpendicular to unit cell direction, respectively.

As an illustrative example, let us consider the $C_6$ rotation (the most nontrivial symmetry here). In general the six monopoles could transform as $\Phi_i\to e^{i\theta_{C_c}}O_{ij}\Phi_j$ with $e^{i\theta_{C_6}}$ an overall phase and the $SO(6)$ matrix $O$ given by
\be
O=\left( \begin{array}{cccccc}
                  -\cos(2\pi/3) &  \sin(2\pi/3) & 0 & 0 & 0 & 0 \\
                  \sin(2\pi/3) & \cos(2\pi/3) & 0 & 0 & 0 & 0 \\
                  0 & 0 & 1 & 0 & 0 & 0 \\
                  0  & 0 & 0 & -1 & 0 & 0 \\
                  0  & 0 & 0 & 0 & -1 & 0 \\ 
                  0  & 0 & 0 & 0 & 0 & -1 \end{array}\right).
\ee
The form of the matrix $O_{ij}$ is fixed by the symmetry transforms of the Dirac fermions $C_6: \psi\to \tau^1\exp\left(-i\frac{2\pi}{3} \tau^3 \right)\psi$, up to an overall sign that can be absorbed into the overall phase factor. Our argument implies that the phase factor $e^{i\theta_{C_6}}$ should be chosen so that the total transform takes the form
\be
\left( \begin{array}{cc} 1 & 0 \\ 0 & SO(5) \end{array} \right),
\ee
with the trivial monopole being a spin singlet. Since $O$ already takes the above form (with $\Phi_3$ being a trivial singlet), the additional $U(1)_{top}$ phase factor must be trivial. Focusing on the spin triplet monopoles ($\Phi_{4,5,6}$), this reproduces the result previously obtained through numerical calculations\cite{ran_2008}. 

The other symmetry actions can be determined using the same logic. The results are tabulated in Table~\ref{table:honeycomb_monopole}.

\begin{widetext}
\begin{table*}
\captionsetup{justification=raggedright}
\begin{center}
\begin{tabular}{|c|c|c|c|c|c|}
\hline
& $T_1$ & $T_2$ & $ R$ & $ C_6$ & $\mathcal T$\\

\hline
$\Phi_1^\dagger$ &  \multicolumn{2}{c|}{$ \cos(\frac{2\pi}{3}) \Phi_1^\dagger-\sin(\frac{2\pi}{3}) \Phi_2^\dagger$} &$\Phi_1$ & $ -\cos(\frac{2\pi}{3}) \Phi_1^\dagger+\sin(\frac{2\pi}{3}) \Phi_2^\dagger$ & $\Phi_1$\\
$\Phi_2^\dagger$ &  \multicolumn{2}{c|}{$ \cos(\frac{2\pi}{3}) \Phi_2^\dagger+\sin(\frac{2\pi}{3}) \Phi_1^\dagger$} & $-\Phi_2$ & $ \cos(\frac{2\pi}{3}) \Phi_2^\dagger+\sin(\frac{2\pi}{3}) \Phi_1^\dagger$& $\Phi_2$\\
$\Phi_3^\dagger$ & $\Phi_3^\dagger$& $\Phi_3^\dagger$&$\Phi_3$ & $\Phi_3^\dagger$ & $\Phi_3$\\
$\Phi_{4/5/6} ^\dagger$ & $\Phi_{4/5/6}^\dagger $& $\Phi_{4/5/6}^\dagger $&  $-\Phi_{4/5/6} $ &$-\Phi_{4/5/6}^\dagger $ & $-\Phi_{4/5/6}$\\
\hline
\end{tabular}
\caption{The transformation of monopoles on honeycomb lattice. $T_{1/2}$ is translation along two basic lattice vectors, $R$ is reflection perpendicular to the axis defined by a unit cell (shown in fig~\ref{fig:honeycomb2}, $C_6$ is the six-fold rotation around the center of a plaquette. There is a trivial monopole, i.e, the third monopole $\Phi_3$. 
}
\label{table:honeycomb_monopole}
\end{center}
\end{table*}
\end{widetext}

An important conclusion is that the operator $\Phi_3+\Phi_3^{\dagger}$ transforms trivially under all physical symmetries, and is therefore allowed as a perturbation to the QED$_3$ theory. 

\subsubsection{Square lattice}
We consider the staggered flux state on the square lattice, with lattice hopping amplitudes (in a certain gauge)
\be
\label{eqn:SFhopping}
t_{i,i+\hat{y}}=\exp[i(-1)^{x+y}\theta/2]t, \hspace{5pt} t_{i,i+\hat{x}}=(-1)^yt.
\ee

The staggered flux state can be continuously tuned (without changing the low-energy Dirac dispersion) to have $\theta=0$, also known as the $\pi$-flux state.
The $\pi$-flux state has all the symmetries of the staggered flux state, together with an additional particle-hole symmetry. In fact the $\pi$-flux state has an $SU(2)$ gauge symmetry and the unit cell contains two sites (sublattice $A,B$) with a vertical link. As far as monopole quantum number is concerned, there is no distinction between the two except for the particle-hole symmetry which does not exist in the staggered flux state. We will therefore calculate the monopole quantum number in the $\pi$-flux state, which is simpler.

There are two gapless points in the reduced Brillouin zone at $\bf Q=(\pi/2,\pi), \bf Q'=-\bf Q$. The low-energy theory takes the standard Dirac form in an apropriate basis.

In the $\pi$-flux phase, we can write the (projective) symmetry realizations on the Dirac fermions as
 \begin{align}
 \label{eqn:square_psg}
T_1&: \psi\rightarrow i\tau^3\psi\quad T_2: \psi\rightarrow i\tau^1\psi\nonumber\\
R_x&: \psi\rightarrow \mu^3\ \tau^3\ \psi\quad C_4: \psi\rightarrow e^{i\frac{\pi}{4}\mu^1}e^{-i\frac{\pi}{4}\tau^2} \psi\nonumber\\
\mathcal T&: \psi\rightarrow i\mu^2 \sigma^2\ \tau^2\   \psi, \quad 
\end{align}
together with a particle-hole symmetry:
\be
\mathcal C: \psi\rightarrow i \mu^3 \sigma^2\   \psi^*,
\ee
where $\mu^i$ are Pauli matrices acting on the Dirac spinor index, and $C_4$ means a four-fold rotation around a lattice site. If we turn on a nonzero $\theta$ in Eq.~\eqref{eqn:SFhopping} and convert the state to the staggered flux phase, $\mathcal{C},T_{1,2}, C_4, \mathcal{T}$ will be broken, but $\mathcal{C}T_{1,2}, \mathcal{C}C_4, \mathcal{C}\mathcal{T}$ will be preserved and provide the realizations of the physical symmetries.

The monopole quantum numbers can now be deduced using the argument outlined above. The results are tabulated in Table~\ref{table:square_monopole}. To be concrete, we start with the $\pi$-flux state (first four rows in Table~\ref{table:square_monopole}). The $C_4$ operation, based on its action on the Dirac fermions, should act on the monopoles as $C_4: \Phi_i\to e^{i\theta_{C_4}}O^{C_4}_{ij}\Phi_j$, where
\be
O^{C_4}=\left( \begin{array}{cccccc}
                  0 &  0 & -1 & 0 & 0 & 0 \\
                  0 & 1 & 0 & 0 & 0 & 0 \\
                  1 & 0 & 0 & 0 & 0 & 0 \\
                  0  & 0 & 0 & 1 & 0 & 0 \\
                  0  & 0 & 0 & 0 & 1 & 0 \\ 
                  0  & 0 & 0 & 0 & 0 & 1 \end{array}\right).
\ee
Based on the argument outlined before, we should have $e^{i\theta_{C_4}}=1$ and $\Phi_2$ is the $SO(5)$ singlet monopole. This in turn fixed the actions of $T_{1,2}$ on monopoles. For example, the action of $T_1$ on the Dirac fermions requires that $T_1: \Phi_i\to e^{i\theta_{T_1}}O^{T_1}_{ij}\Phi_j$ where
\be
O^{T_1}=\left( \begin{array}{cccccc}
                  -1 &  0 & 0 & 0 & 0 & 0 \\
                  0 & -1 & 0 & 0 & 0 & 0 \\
                  0 & 0 & 1 & 0 & 0 & 0 \\
                  0  & 0 & 0 & 1 & 0 & 0 \\
                  0  & 0 & 0 & 0 & 1 & 0 \\ 
                  0  & 0 & 0 & 0 & 0 & 1 \end{array}\right).
\ee
Since $\Phi_2$ should be a singlet under flavor symmetries, we must have $e^{i\theta_{T_1}=-1}$, which gives the transformation tabulated in Table~\ref{table:square_monopole}.

The combined symmetry action $\mathcal{CR}_x$, as we discussed before, can act on the monopoles with potentially nontrivial Berry phases. For this purpose we can view this symmetry as a combination of $\mathcal{CR}_0$ (as defined in Eq.~\eqref{eqn:C0} and \eqref{eqn:R0}) and a flavor rotation $\psi\to i\tau^1\psi$, followed by a Lorentz rotation which does not affect the scalar monopoles. The $\mathcal{CR}_0$ transforms the monopoles as Eq.~\eqref{T0CR0} and the flavor rotation is essentially the $T_2$ transformation. Therefore under $\mathcal{CR}_x$ we should have $\Phi_1\to -\Phi_1$ and $\Phi_i\to\Phi_i$ for $i\neq1$. For $\mathcal{R}_x$ and $\mathcal{C}$ separately, the overall phase depends on the definition of the monopoles as we discussed, but the relative transformation between the six monopoles is still meaningful. Since $\mathcal{R}_x$ involves a flavor rotation, it should act on the monopoles as shown in Table~\ref{table:square_monopole} up to an overall phase which we fix to be trivial. The transformation of $\mathcal{C}$ then follows immediately. Finally the transformation of $\mathcal{T}$ on the monopoles is simply given by Eq.~\eqref{T0CR0}.

\begin{table*}
\captionsetup{justification=raggedright}
\begin{center}
\begin{tabular}{|c|c|c|c|c|c|c|}
\hline
& $T_1$ & $T_2$ & $ R_x$ & $ C_4$ &$\mathcal C$& $\mathcal T$\\

\hline
$\Phi_1^\dagger$ &  $+$ & $-$ &$-\Phi_1$& $-\Phi_3^\dagger$&$\Phi_1$&$\Phi_1$\\
$\Phi_2^\dagger$ & $+$ & $+$ &$-\Phi_2$& $\Phi_2^\dagger$&$-\Phi_2$&$\Phi_2$\\
$\Phi_3^\dagger$ & $-$ & $+$ &$\Phi_3$& $\Phi_1^\dagger$&$\Phi_3$ &$\Phi_3$ \\
$\Phi_{4/5/6} ^\dagger$ &$-$ & $-$ &$\Phi_{4/5/6}$& $\Phi_{4/5/6}^{\dagger}$ &$\Phi_{4/5/6}$ &$-\Phi_{4/5/6}$ \\
\hline
$\Phi_1^\dagger$ &  $\Phi_1$ & $-\Phi_1$ &$-\Phi_1$& $-\Phi_3$&  &$\Phi_1^{\dagger}$ \\
$\Phi_2^\dagger$ & $-\Phi_2$ & $-\Phi_2$ &$-\Phi_2$& $-\Phi_2$&  &$-\Phi_2^{\dagger}$\\
$\Phi_3^\dagger$ & $-\Phi_3$ & $\Phi_3$ &$\Phi_3$& $\Phi_1$&  &$\Phi_3^{\dagger}$ \\
$\Phi_{4/5/6} ^\dagger$ &$-\Phi_{4/5/6}$ & $-\Phi_{4/5/6}$ &$\Phi_{4/5/6}$& $\Phi_{4/5/6}$&  &$-\Phi_{4/5/6}^{\dagger}$ \\
\hline
\end{tabular}
\caption{The transformation of monopoles on square lattice. $T_{1/2}$ is translation along two basic lattice vectors, $R_x$ is reflection perpendicular to the horizontal axis, $C_4$ is the four-fold rotation around the site (shown in fig~\ref{fig:honeycomb2}).  There is a trivial monopole by our definition, i.e, the second monopole $\Phi_2$. 
 The first 4 rows and last 4 rows of monopole transformation correspond to $\pi$ flux state and staggered flux state, respectively, which differ by a charge conjugation for $T_{1/2},C_4$ and $\mathcal{T}$. The last 4 rows align with the results of $M$ transformations of Ref \onlinecite{alicea_2008} after making the identification $\Phi_{1/3}=M_{3/2},\Phi_2=iM_1,\Phi_4\mp i\Phi_5=M_{4/6},\Phi_6=M_5$ where $M^\dagger$'s are the ``monopole operators" defined in Ref.~\onlinecite{alicea_2008}. We emphasize that the ``$\pi$-flux state" with $U(1)$ gauge symmetry does not actually represent a spin system, and should be viewed as an intermediate state for our purpose.}
 
\label{table:square_monopole}
\end{center}
\end{table*}

Now we proceed to consider the staggered-flux state (the actual $U(1)$ Dirac spin liquid). The only difference now is that $\mathcal{C}$ is no longer a symmetry, and the action of $T_{1,2}, C_4, \mathcal{T}$ should all be combined with $\mathcal{C}$. The results are tabulated in Table~\ref{table:square_monopole} in the last four rows. Again we see that there is a trivial monopole ${\rm Im}\Phi_2$ that can be added as a perturbation.

\subsection{Another approach: lattice symmetries generated by $\mathcal{CR}$}

There is another trick to obtain lattice symmetry quantum number for the monopoles on bipartite lattices, thanks to the extra $\mathcal{C}$ symmetry. The key is to realize that rotation and translation symmetries can all be generated by repeatedly applying $\mathcal{CR}$ symmetries with respect to different reflection axes (simple $\mathcal{R}$ will not work since the overall phase is not well defined). We will show below that this approach gives results identical to those obtained in Sec.~\ref{MQNQCD}, which gives us more confidence since the two approaches appear to be very different from each other.

Following Sec.~\ref{dimred}, we calculate the monopole lattice quantum numbers when the fermions form a quantum spin Hall insulator (which preserves all lattice symmetries). Recall that in the quantum spin Hall phase, $\mathcal{CR}: \mathcal{M}\to \pm\mathcal{M}$ if $\mathcal{CR}^2=\mp1$ on the fermions. This gives us the quantum numbers of a particular monopole (say $\Phi_4+i\Phi_5$), and those of other monopoles can be fixed by the $SO(6)$ flavor symmetry. Therefore our task below is to show that the quantum numbers of $\Phi_4+i\Phi_5$ calculated in the quantum spin Hall phase agree with those tabulated in Table~\ref{table:honeycomb_monopole} and \ref{table:square_monopole}.

\subsubsection{Honeycomb}
\begin{center}
\begin{figure}

\captionsetup{justification=raggedright}

\adjustbox{trim={0\width} {0\height} {0\width} {0\height},clip}
{\includegraphics[width=1\columnwidth]{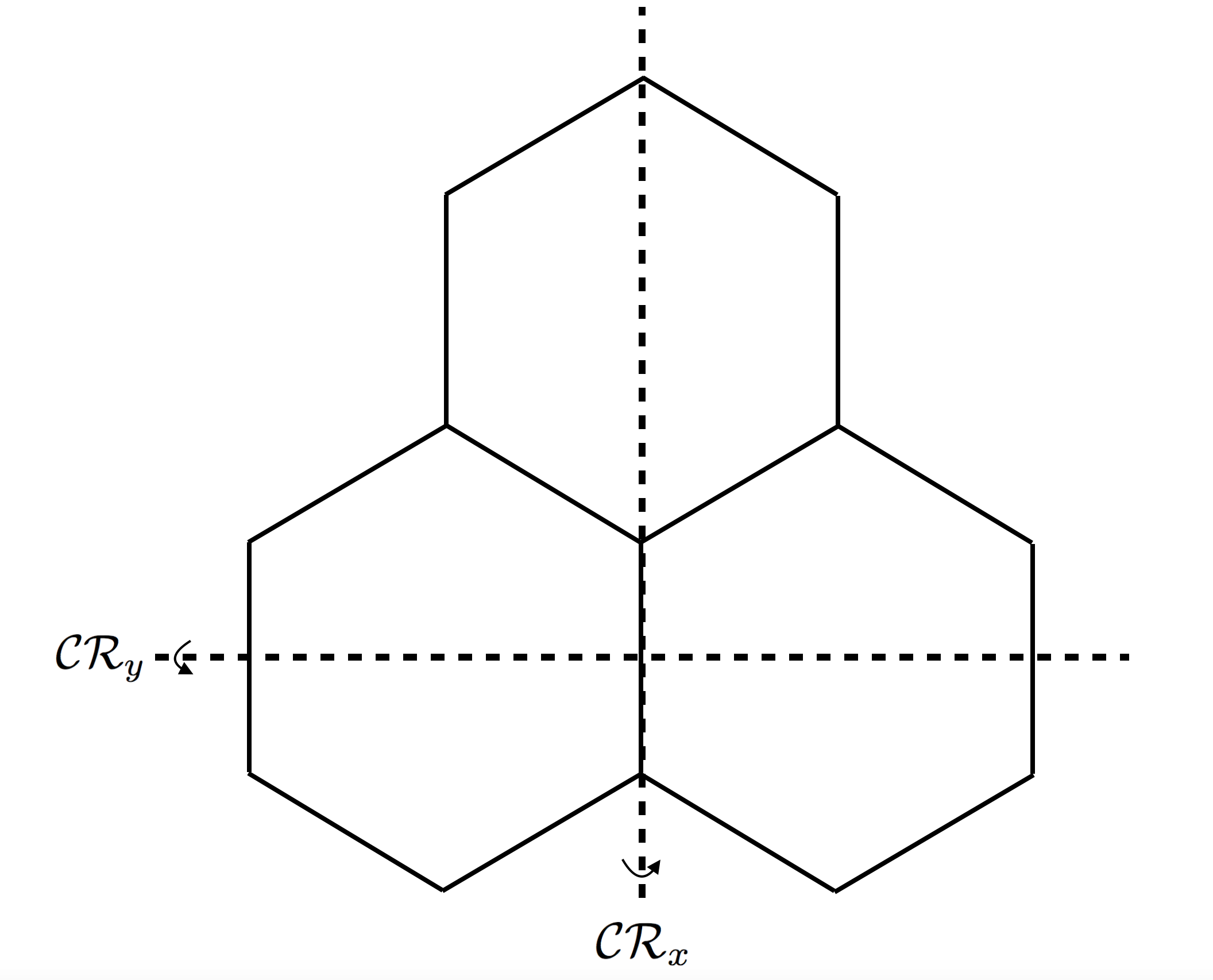}}

\caption{Reflection axis to be considered in the main text.}

\label{fig:Honeycombreflection}
\end{figure}
\end{center}

Here we find $(\mathcal C R_{x/y})^2=-1 (1)$ when acting on fermions, (see Fig.~\ref{fig:Honeycombreflection}). This means that (from Sec.~\ref{dimred}) the spin triplet monopoles ($\Phi_{4,5,6}$) stays invariant/ gets a minus sign under $\mathcal C R_x, \mathcal CR_y$, respectively. All other space symmetries are generated from reflections and we get the transformation results in table \ref{table:honeycomb_monopole}. For example, $C_6$ operation can be obtained by two different $\mathcal{CR}$ reflections in succession, and we immediately see that $C_6: \Phi_{4,5,6}\to \Phi_{4,5,6}$. Likewise, using two different reflections to generate translations leads to the results $T_{1,2}: \Phi_{4,5,6}\to \Phi_{4,5,6}$. These are all in agreement with Table~\ref{table:honeycomb_monopole}.

We emphasize here that the above logic works because if two different reflection axes are related by a symmetry operation, then these two reflections should act on the (unique) monopole $\mathcal{M}$ in the same way. More precisely, $g(\mathcal{CR}_1)g^{-1}=\mathcal{CR}_1$ when acted on an one-dimensional representation.

\subsubsection{Square}

\begin{center}
\begin{figure}

\captionsetup{justification=raggedright}

\adjustbox{trim={.18\width} {.05\height} {0\width} {.05\height},clip}
{\includegraphics[width=1.2\columnwidth]{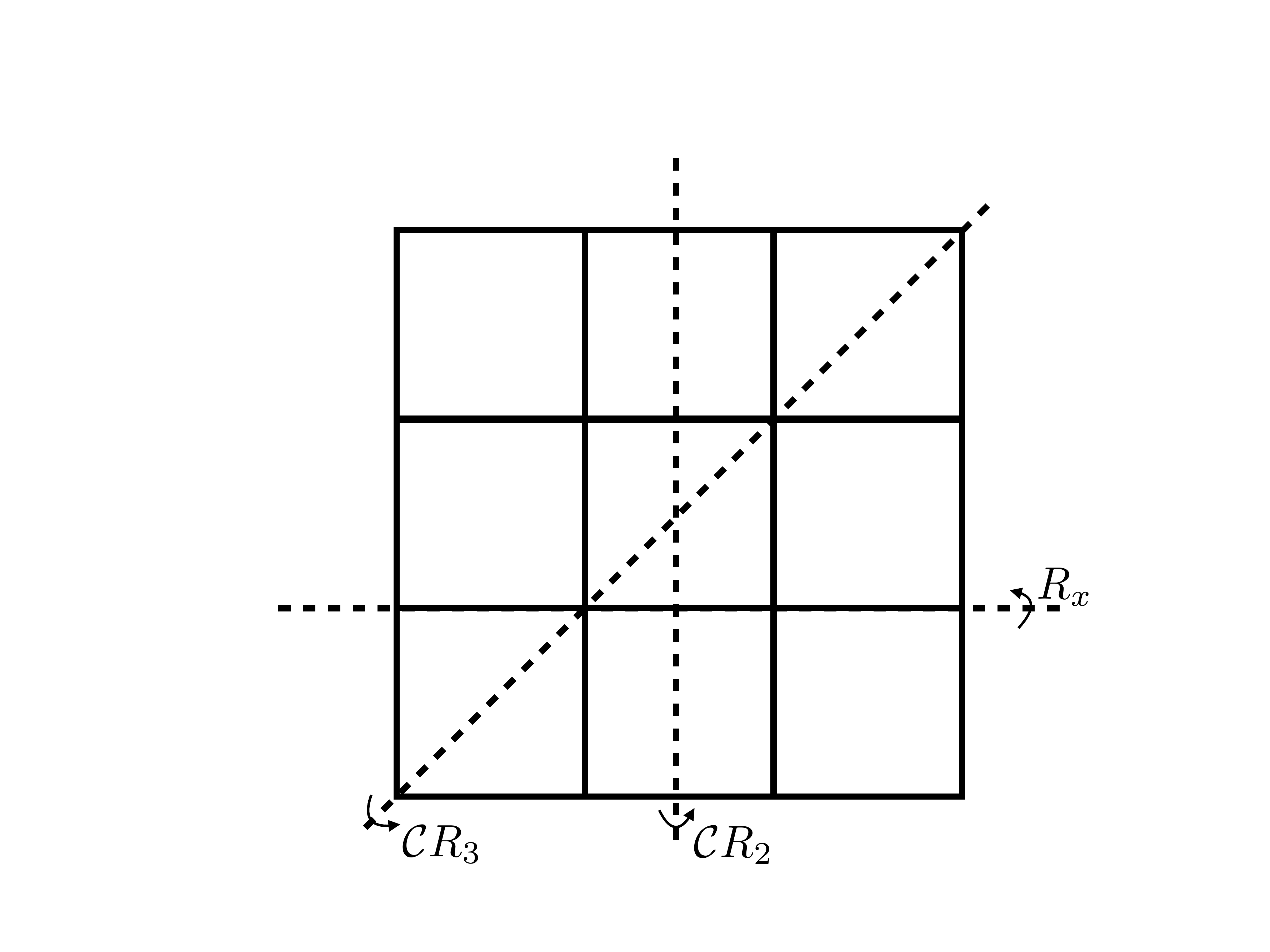}}

\caption{Reflection axis to be considered in the main text.}

\label{fig:reflections}
\end{figure}
\end{center}

As discussed before, it suffices to consider the $\pi$-flux phase which has a charge conjugation symmetry. Now consider $\mathcal{CR}_1\equiv\mathcal{C}R_x$, $\mathcal{C}R_2\equiv \mathcal C T_1 C_4^2R_x$ and $\mathcal{C}R_3\equiv \mathcal C C_4T_1C_4^2R_x$ -- the last two are reflections across axis labeled in Fig.~\ref{fig:reflections}.

It is easy to check that $(\mathcal{C}R_1)^2=(\mathcal{C}R_3)^2=-1$ and $(\mathcal{C}R_2)^2=+1$ on the fermions, which immediately leads to
\bea
\mathcal{CR}_1:&& \mathcal{M}\to \mathcal{M} \nn
\mathcal{CR}_2:&& \mathcal{M}\to -\mathcal{M} \nn
\mathcal{CR}_3:&& \mathcal{M}\to \mathcal{M}.
\eea

It is now straightforward to read off other symmetry transformations on this monopole $\mathcal{M}\sim \Phi_4+i\Phi_5$. For example $C_4\sim \mathcal{CR}_3\cdot \mathcal{CR}_1: \mathcal{M}\to \mathcal{M}$ and $T_{1,2}\sim \mathcal{CR}_1\cdot\mathcal{CR}_2: \mathcal{M}\to-\mathcal{M}$. 

We also make a phase choice so that
\be
\mathcal{C}: \mathcal{M}\to \mathcal{M}^{\dagger}.
\ee
The transformation of charge conjugation $\mathcal C$ depends on the phase choice of monopole but the relative phase between $6$ monopoles are fixed and hence meaningful.

Then we essentially have reproduced Table~\ref{table:square_monopole} for the $\pi$-flux state. The transformation for $\Phi_{1,2,3}$ can be fixed by further applying the emergent $SO(6)$ symmetry.

\section{A more general scheme: atomic (Wannier) centers}
\label{wanniercenter}
 We have seen that on bipartite lattices (honeycomb and square), with the help of particle-hole symmetry (a hallmark of bipartite lattices), the monopole quantum numbers under lattice rotation and translation can be uniquely fixed. Here we shall discuss a method applicable on all lattices including non-bipartite ones such as triangular and kagom\'e. 

Notice that in this problem lattice rotation plays a more fundamental role than lattice translation in two ways. First, if lattice rotation symmetry is absent, there will be no reason for the monopole to have a quantized lattice momentum, which can take continuous value in the entire Brillouin zone. Now if we impose certain rotation symmetry $C_n$ ($n=2,3,4,6$), the momentum will only take certain (discrete) rotationally invariant values, which will be robust as the system is adiabatically deformed. Second, once we know how the monopoles transform under rotations around different centers, we know automatically how it should transform under translation since translation operation can be generated by subsequent rotations around different centers.

Presumably the lattice momentum and angular momentum of the monopole operator is also decided by the ``band topology" of the spinon insulator, just like time-reversal symmetry -- but how? Before answering this question systematically, let us consider a more familiar example of bosonic spinon (Schwinger boson) theory, where it is well known that on bipartite lattices the monopole carries $l=\pm1$ angular momentum under site-centered rotations, leading to valence bond solid (VBS) order when the spinons are gapped\cite{HaldaneBerry, ReSaSUN} -- a fact important in the context of deconfined quantum phase transition between Neel and VBS states\cite{senthil_20031,senthil_20041}. We now review this fact with a physically intuitive picture, which could then be generalized to more complicated cases with fermionic spinons. For the fermionic spinons, we will illustrate the new method in this section with the bipartite examples. We will obtain the same results as in Sec.~\ref{MQNBipartite} with considerably more involved calculations -- the goal being to establish the method in a setting where the answers are independently known, so that we can apply it to the DSL on non-bipartite lattices (and further confirm the results on bipartite lattices).

\subsection{ Warm-up: bosonic spinons and valence bond solid}
\label{warmup}

For concreteness let us consider a honeycomb lattice with spin-$1/2$ per site. In the Schwinger boson formulation we decompose the spin operator as
\be
\vec{S}_i=(-1)^i\frac{1}{2}b^{\dagger}_{i,\alpha}\vec{\sigma}_{\alpha\beta}b_{i,\beta},
\ee
where $i$ is the site label and $(-1)^i$ takes $\pm1$ on the two sub-lattices, respectively. $b_{\alpha}$ is a hard-core boson ($(b^{\dagger}_{\alpha})^2$=0) with spin $\alpha\in\{\uparrow,\downarrow\}$. The physical Hilbert space has $\sum_{\alpha}b^{\dagger}_{\alpha}b_{\alpha}=1$. Similar to the fermionic spinon theory, this constraint only needs to be satisfied on average in a spin liquid phase. There is again a dynamical $U(1)$ gauge field $a_{\mu}$ coupled to the Schwinger bosons, with gauge charge on each site defined as $q_i=\sum_{\alpha}b^{\dagger}_{\alpha}b_{\alpha}-1$.

The $(-1)^i$ factor in the parton decomposition is chosen so that when the Schwinger bosons condense in a uniform manner, the spins order as a collinear anti-ferromagnet (Neel state). Due to this $(-1)^i$ factor, under the $C_6$ rotation (which exchanges the two sub-lattices), the Schwinger bosons should transform (in addition to the coordinate change) as
\be
b_{\alpha}\to i\sigma^y_{\alpha\beta}b^{\dagger}_{\beta}.
\ee
Similar transform also happens on square lattice under translation. 

\begin{center}
\begin{figure}

\captionsetup{justification=raggedright}

\adjustbox{trim={.18\width} {.15\height} {0\width} {.15\height},clip}
{\includegraphics[width=1.5\columnwidth]{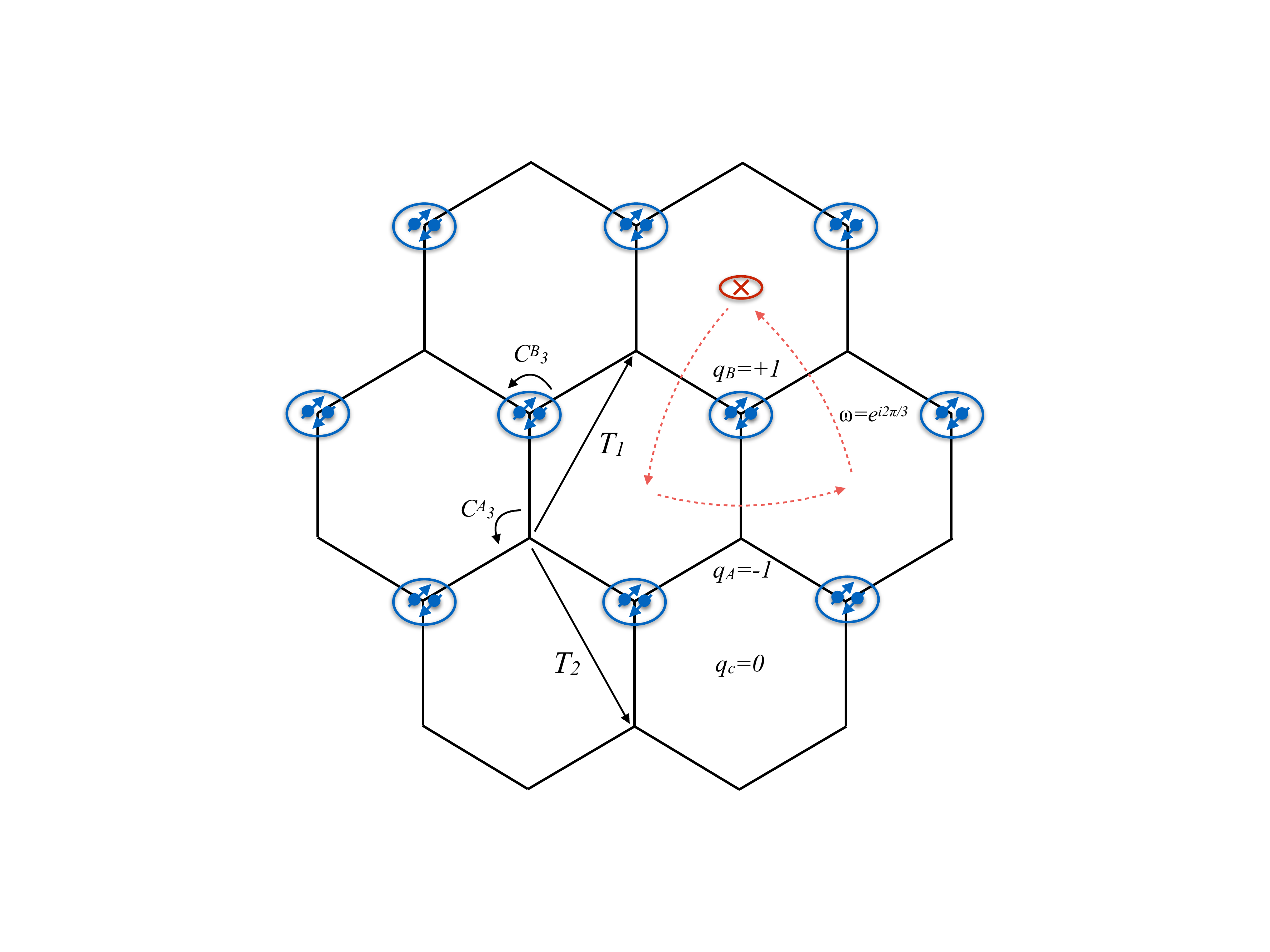}}

\caption{The simplest spinon (boson) Mott insulator, with every site in A sub-lattice empty, and every site in B sub-lattice completely filled (recall that the bosons $b_{\alpha}$ are hard-core). This respects $C_6$ rotation since it acts as $b_{\alpha}\to i\sigma^y_{\alpha\beta}b^{\dagger}_{\beta}$. As a monopole (flux) moves in this background, it sees a fixed gauge charge pattern with $q_A=-1$ and $q_B=+1$. The amount of gauge charge sitting on each rotation center dictates the nontrivial Berry phase accumulated by the monopole as it moves around the center according to $\theta(C_n^r)=e^{iq_r2\pi/n}$. The translation quantum numbers of the monopole can be easily obtained once the rotation quantum number is known.}

\label{fig:honeycomb1}
\end{figure}
\end{center}

We would like to construct a state in which the Schwinger bosons are gapped, i.e. they form a bosonic Mott insulator. The simplest such state respecting all the symmetries -- especially spin rotation and $C_6$ rotation -- is shown pictorially in Fig.~\ref{fig:honeycomb1}. In this state every site in A sub-lattice is empty, and every site in B sub-lattice has both bosonic orbits occupied. The wavefunction is simply a product state
\be
\prod_{i\in A}|0\rangle_i\ \prod_{j\in B}b^{\dagger}_{j, \uparrow}b^{\dagger}_{j,\downarrow}|0\rangle_j.
\ee

Now what happens when a monopole (a flux quanta) moves in this simple charge background? What the flux sees is a fixed gauge charge pattern, with $q_A=-1$ on each A-site and $q_B=+1$ on each B-site. Therefore as the flux moves around each site, a non-trivial Berry phase is picked up. The amount of gauge charge $q_r$ sitting on each rotation center $r$ dictates the Berry phase under the $C_n$ rotation to be 
\be
\label{eqn:berryphaseformula}
\theta(C_n^r)=e^{iq_r2\pi/n}.
\ee
Effectively the monopole gains an angular momentum $l_A=-1$ under A-site-centered $C^A_3$ rotation (corresponding to a Berry phase $\omega^{-1}=e^{-i2\pi/3}$ under a $C^A_3$ rotation), and $l_B=+1$ under B-site-centered $C^B_3$ rotation (Berry phase $\omega$), and $l_c=0$ under plaquette-centered $C^c_6$ rotation since there is no charge placed at the center of the hexagon plaquettes. One may ask how robust these Berry phases are -- for example, will the results change if the flux is inserted far away from the rotation center? The answer is no, since by $C_n$ symmetry the total charge enclosed by a closed rotation trajectory will always be $q_r$ (mod $n$). Therefore the monopole Berry phase will always be given by Eq.~\eqref{eqn:berryphaseformula} under a $C_n$ rotation. This is essentially the spirit of dimensional reduction introduced in \cite{song_2017}.

Translation symmetry quantum numbers are now easily obtained: $T_1=C_A^{-1}C_B=\omega^2$ and $T_2=C_BC_A^{-1}=\omega^2$. This makes the monopole operator the Kakule VBS order parameter, as expected.

The above argument can be extended straightforwardly to square lattice, from which the standard results follow, namely the monopole carries $\pm1$ angular momentum under site-centered $C_4$ rotations and is therefore identified with columnar VBS order parameter\cite{HaldaneBerry, ReSaSUN}.

\subsection{Wannier centers on honeycomb lattice: a case study}
\label{Wannier}

\begin{center}
\begin{figure}

\captionsetup{justification=raggedright}

\adjustbox{trim={.10\width} {.22\height} {.10\width} {.02\height},clip}
{\includegraphics[width=1.5\columnwidth]{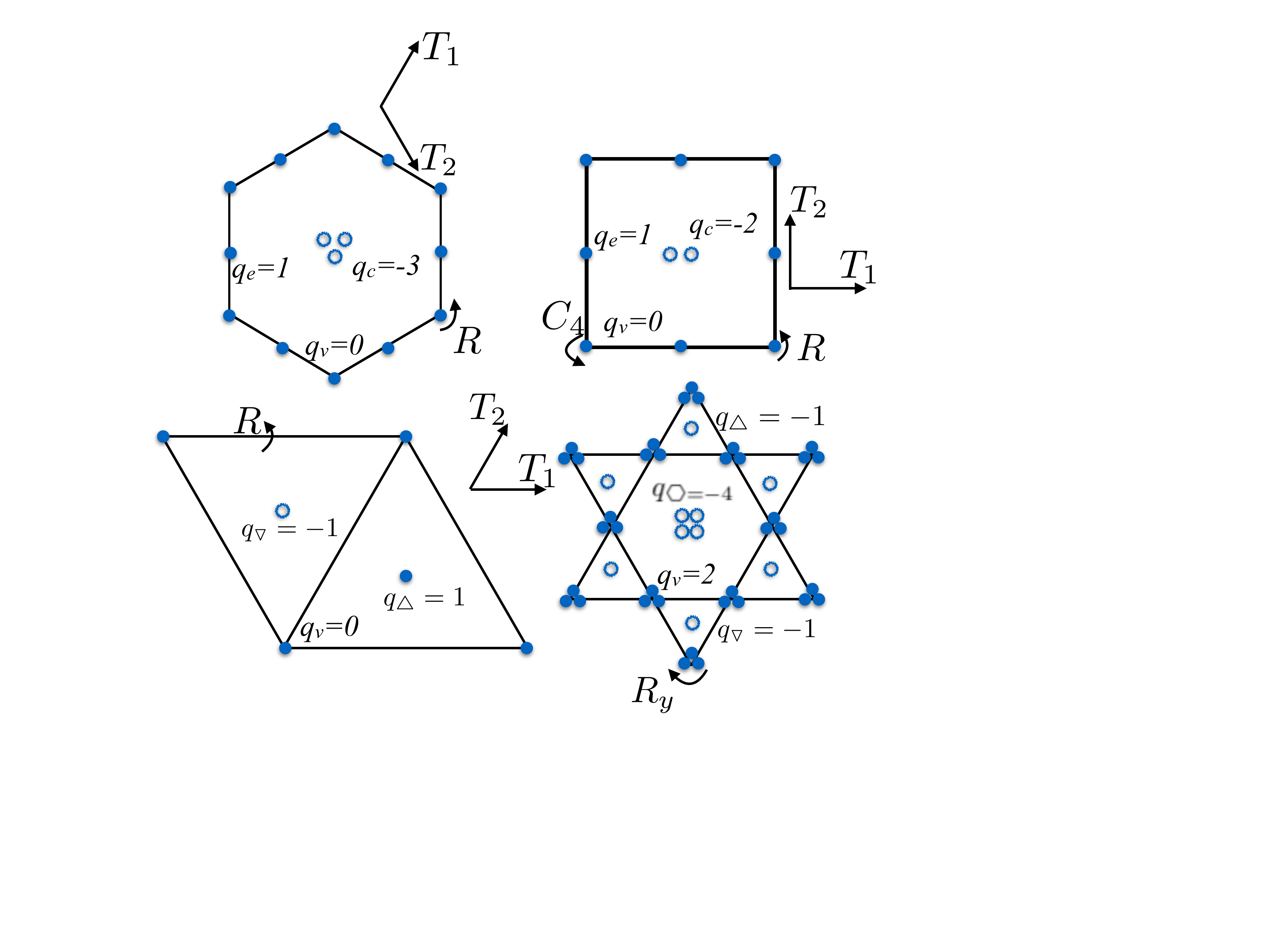}}

\caption{The Wannier limit of a quantum spin Hall insulator on 4 types of lattices when all the symmetries are broken except for charge $U(1)$ conservation and lattice translation/rotation. The solid dot indicates a Wannier center and empty dot indicates ``minus" charge pertinent to the fragile nature of the band topology. 
The pattern of $U(1)$ gauge charge is shown in the figure, with the background $-1$ charge per site included.}

\label{fig:honeycomb2}
\end{figure}
\end{center}

We now return to fermionic spinons in a quantum spin Hall insulator band structure. Let us first consider the simpler honeycomb lattice. The band structure is given by the Haldane model\cite{HaldaneHoneycomb} for each spin component with opposite Chern numbers. The hopping amplitudes include a real nearest-neighbor hopping and an imaginary second-neighbor hopping that preserve all the lattice symmetries and time-reversal symmetry $f\to i\sigma^2 f$. One may ask whether we can deform this insulator to an ``atomic limit" as we did above in Sec.~\ref{warmup}, and then trivially read off the monopole's lattice quantum numbers. Obviously this is impossible if time-reversal and spin rotation (now only $SO(2)$) symmetries are preserved, since the spinons form a strong topological insulator and therefore (almost by definition) an atomic limit does not exist. However, since the question of monopole lattice quantum number has nothing to do with spin rotation and time-reversal symmetry, we may as well explicitly break all the symmetries except the lattice translation, rotation, and the $U(1)$ charge conservation. In the simplest setting, the resulting insulator can be deformed into an ``atomic limit", which can be pictured as particles completely localized in real space.
One nontrivial aspect, compared to the bosonic spinon case in Sec.~\ref{warmup}, is that the effective centers of the localized orbits do not have to sit on the original lattice sites (where the microscopic fermions are defined). For insulators described by free fermion band theory these are simply the Wannier centers. One can always deform all the Wannier centers, without further breaking any symmetry, to one of the rotation centers, such as lattice sites or plaquette centers (see Fig.~\ref{fig:honeycomb2} for example). Once such a configuration is reached, we can simply follow the procedure in Sec.~\ref{warmup} to obtain the monopole Berry phase under a $C_n$ rotation centered at $r$ according to Eq.~\eqref{eqn:berryphaseformula}. Lattice momentum of the monopole can then be obtained by composing rotations at different centers.

Let us illustrate this with the QSH insulator on honeycomb lattice -- for concreteness we assume that the nearest-neighbor hopping $t>0$. To determine the exact nature of the Wannier limit, we employ the techniques developed in Refs.~\cite{PoIndicators,Pofragile,Cano_2018}. The basic idea is to look at high symmetry points in momentum space, namely momentum points that are invariant under various lattice rotations. The fermion Bloch states at these high symmetry points will now form certain representations of the rotation symmetries, and our task is to determine what kind of atomic (Wannier) limit would give rise to such a representation.

Consider inversion symmetries $I^c_2$ with respect to a hexagon center. There are four inversion invariant momentum points: $\Gamma$ ($T_1=T_2=1$), $M$ ($T_1=T_2=-1$), and two other points $M'$, $M''$ related to $M$ by $C_6$ rotations. Since there are in total two bands occupied (one per spin), the Bloch states form a two dimensional reducible representation of $I^c_2$ at each of these momentum point. The spectrum of inversion eigenvalues $\{\lambda_{I^c}\}$ at each momentum is
\bea
\label{eqn:hexagoninversion}
\lambda_{I^c}^{\Gamma}&:& \{+1, +1\}, \nn
\lambda_{I^c}^{M}&:& \{-1, -1\}.
\eea
Now what kind of atomic insulator can form such representations at $\Gamma$ and $M$? Let us consider several candidates. First, consider the simplest insulator, with exactly one Wannier center sitting on each lattice site. It is easy to show that this insulator has $\lambda_{I^c}: \{+1,-1\}$ at both $\Gamma$ and $M$. Next consider an insulator with one Wannier center sitting at the center of each edge. This has $\lambda_{I^c}^{\Gamma}: \lambda_0\times\{1,1,1\}$ and $\lambda_{I^c}^{M}: \lambda_0\times\{1,-1,-1\}$ where $\lambda_0$ is the intrinsic inversion eigenvalue of the Wannier orbit which can be either $\pm1$. There is yet another insulator with one Wannier center sitting at the center of each hexagon plaquette. This has $\lambda_{I^c}^{\Gamma}=\lambda_{I^c}^M=\lambda_0$ for some intrinsic $\lambda_0=\pm1$. We have exhausted all distinct types of atomic (Wannier) insulators but it appears that none of these fit the representation of our occupied band Eq.~\eqref{eqn:hexagoninversion}! The key concept we need here is that of fragile topology\cite{Pofragile}: the occupied band, even though being ``topologically trivial", can be deformed into an atomic (Wannier) limit only when combined with another atomic insulator. In our case, if we combine the occupied band with an atomic insulator with three Wannier centers sitting on each hexagon center, with intrinsic inversion eigenvalues $\lambda_0=\{1,1,-1\}$, the total system will form inversion representations at $\Gamma$ and $M$ that resembles an atomic insulator with one Wannier center on each site and at the center of each edge (with $\lambda_0=1$). Formally we can write this relation as (upper left of Fig \ref{fig:honeycomb2})
\bea
honeycomb\ QSH\ &&occupied\ band=site+edge^{\lambda_0=1} \nn &&-2\times hexagon^{\lambda_0=1}-hexagon^{\lambda_0=-1},
\eea
where the minus signs indicate the fragile nature of the topology. From the monopole point of view, the bands that are formally subtracted in the above relation can simply be viewed as negative gauge charges sitting on their Wannier centers. So the above relation produces a gauge charge pattern shown in Fig.~\ref{fig:honeycomb2} (remember that there is always a $-1$ background gauge charge sitting on each site). 

One should also check that the above Wannier pattern also reproduces representations at high symmetry points of other point group symmetries such as site-centered $C_3$. Since the $C_3$-invariant momentum points include the Dirac points ($Q$ and $Q''$), it is important for us to include the Dirac mass term that corresponds to the QSH mass. It is a relatively straightforward exercise to check that the above Wannier pattern indeed reproduces the representations from the occupied band of the QSH insulator.

The gauge charge pattern thus produced gives rise to some nontrivial quantum numbers for the monopole. Using Eq.~\eqref{eqn:berryphaseformula}, we conclude that the monopole transforms under $C_6$ as $\mathcal{M}\to -\mathcal{M}$, and transforms trivially under lattice translations. This is exactly what we found in Sec.~\ref{MQNBipartite} and numerically found in Ref.~\onlinecite{ran_2008}.

We comment that we can alternatively focus on the un-occupied bands of the QSH insulator (holes) instead of the occupied bands (particles), and ask about Wannier centers of the holes. Since the occupied and un-occupied bands together produce two trivial Wannier orbits on each lattice site, it can be shown straightforwardly that we would obtain the identical gauge charge pattern by considering holes instead of particles.

\subsection{Wannier centers on Square lattice}
\label{square}

Now consider the $\pi$-flux state on square lattice. We deform the band structure to the QSH regime. One could obtain how monopoles transform under various rotations by deforming the spinon bands to a Wannier insulator as we did above on honeycomb.
One finds that on square lattice (upper right of Fig \ref{fig:honeycomb2})
\bea
square\ QSH\ &&occupied\ band=site\nn &&+edge\ center -2\times plaquette,
\eea
where the RHS means Wannier insulators with particles lying on site, edge center or plaquette centers. Accounting for the background $1$ per site minus gauge charge, this configuration means the gauge charge on-site vanishes, leading to spin triplet monopoles staying invariant, under site-centered $C_4$ in Table~\ref{table:square_monopole}, while for $C_4$ around plaquette center the monopoles get $-1$ since they seem two gauge charges sitting at the rotation center. This implies that, for example, under $T_1\cdot T_2$ the spin-triplet monopoles should change sign, in agreement with Table~\ref{table:square_monopole}.

We note that similar results were obtained in Ref.~\cite{thomson_2018} by considering a VBS insulator formed by the spinons, in which the effective Wannier centers are simply sitting on the lattice sites and the computation was simpler.

\section{Monopole quantum numbers II: non-bipartite lattices}
\label{calculation}

We are finally ready to calculate monopole lattice quantum numbers on the  non-bipartite lattices i.e.  triangular and kagom\'e lattices.

\subsection{Triangular lattice: Wannier centers from projective symmetry group}
\label{triangular_calculation}

\begin{table}
\captionsetup{justification=raggedright}
\begin{tabular}{|c| c| c| c |c|  c|}
\hline
 &  $T_1$& $T_2$& $R$ &$C_6$ &$\mathcal T$\\
\hline
$\Phi_1^\dagger$ &$e^{i\frac{-\pi}{3}} \Phi_1^\dagger$&$e^{i\frac{\pi}{3}} \Phi_1^\dagger$&$-\Phi_3^\dagger$ &$\Phi_2$&  $\Phi_1$ \\
$\Phi_2^\dagger$ &$e^{i\frac{2\pi}{3}} \Phi_2^\dagger$&$e^{i\frac{\pi}{3}} \Phi_2^\dagger$&$\Phi_2^\dagger$ &$-\Phi_3$ & $\Phi_2$\\
$\Phi_3^\dagger$ &$e^{i\frac{-\pi}{3}} \Phi_3^\dagger$&$e^{i\frac{-2\pi}{3}} \Phi_3^\dagger$&$-\Phi_1^\dagger$ &$-\Phi_1$& $\Phi_3$ \\
$\Phi_{4/5/6} ^\dagger$& $e^{i\frac{2\pi}{3}}\Phi_{4/5/6}^\dagger$&$e^{i\frac{-2\pi}{3}}\Phi_{4/5/6}^\dagger$&$\Phi_{4/5/6}^\dagger$&$-\Phi_{4/5/6}$ & -$\Phi_{4/5/6}$ \\
\hline
\end{tabular}

\caption  {
Monopole transformation laws on triangular lattice. $C_6$ is $6-$ fold rotation around site and other symmetries are marked in figure~\ref{fig:honeycomb2}. There is nontrivial Berry phase for translations.} \label{table:triangular_monopole}
\end{table}

\subsubsection{ Mean-field and PSG}

On triangular lattice we focus on a particular mean-field ansatz with the Hamiltonian
\begin{equation}
\mathcal H=J\sum_{\bra ij\ket} t_{ij} \sum_\alpha f^\dagger_{i\alpha}f_{j\alpha}+h.c.
\end{equation}
where $t_{ij}=\pm 1$ and there is a ``staggered $\pi$ flux" configuration of $t_{ij}$ on the triangular lattice, with a $\pi$-flux on each upward triangle and zero flux on each downward triangle. More details can be found in Appendix~\ref{bilinears}. Under an appropriate basis the low-energy Hamiltonian takes the standard Dirac form, with two spins (denoted by Pauli matrices $\sigma$), two valleys (denoted by Pauli matrices $\tau$). The projective symmetry representation of the Dirac fermions (translation $T_{1/2}$, reflection $R_x$, six-fold rotation $C_6$ and time-reversal) are calculated in standard method and the results are listed below:
\begin{eqnarray}
\psi&&\xrightarrow{T_1}i\tau^3\psi, \nn
\psi&& \xrightarrow{T_2}-i\tau^2\psi, \nn
\psi&& \xrightarrow{\mathcal T}i\sigma^2 \mu^2\tau^2\psi(-k), \nn
\psi&&\xrightarrow{C_6}i\sigma^2  W_{C_6} \psi^\dagger \nn
\psi&&\xrightarrow{R}i\sigma^y  W_{R} \psi^\dagger
\end{eqnarray}
where we have suppressed the coordinate transforms and
\begin{eqnarray}
W_{C_6}&&=e^{-i\gamma^0 \frac{\pi}{6}} W_c \sigma^2 e^{i\frac{\pi}{3}\tau^C}\quad \tau^C=\frac{1}{\sqrt{3}}(\tau^1+\tau^2+\tau^3)\nonumber \nn
W_R&&=\frac{(\gamma^1-\sqrt{3}\gamma^2)}{2} W_c \frac{\tau^3-\tau^1}{\sqrt{2}} \nn
W_c&&=\frac{1}{\sqrt{3}}(-iI_{4\times4}-\mu^3+\mu^1)
\end{eqnarray}
where $\mu^i$ are Pauli matrices acting on the Dirac spinor index (i.e. superposition of $\gamma$ matrices). Again more details can be found in Appendix~\ref{bilinears}.

The above transformations on Dirac fermions fix the monopole transformations up to overall $U(1)_{top}$ phase factors. 
The results are listed in Table~\ref{table:triangular_monopole}. Let us provide some more explanations here: 

  \begin{itemize}
 
 \item {Translations $T_{1/2}$: 
$\psi\xrightarrow{T_1}-i\tau^2\psi,
\psi\xrightarrow{T_2}i\tau^3\psi$,  which gives a relative minus sign to $\Phi_{1/3},\Phi_{1/2}$, respectively. The overall phase factor is undetermined, but is constrained by the invariance of $\Phi_{4/5/6}$ under $C_3=C_6^2$ to take value in $\{1,\omega\equiv e^{i\frac{2\pi}{3}}, \omega^{-1}\}$.}

\item{ Time reversal $\mathcal T$: 
$\psi\xrightarrow{\mathcal T}i\sigma^2 \mu^2\tau^2\psi=\mathcal T_0 \psi$. Then from Sec.~\ref{generalities} we know that $\Phi\rightarrow O_T \Phi^\dagger$, which fixes the overall phase factor.}

\item{Six-fold rotation around site $C_6$:
\be
\psi\xrightarrow{C_6}e^{-i\gamma_0 \frac{\pi}{6}}\{i\sigma^2 W_c \tau^2\}\nonumber\\
exp[i\frac{\pi}{3}\tau^C] \psi^*,
\ee
 where the part inside $\{\}$ is identical to  $\mathcal C_0$ defined in eq~\eqref{eqn:C0}}. We disregard the first Lorentz rotation since the monopole is a scalar anyway. The last part involves certain $SO(3)$ rotation in $\Phi_{1/2/3}$ induced by $\tau^2 [\frac{1}{2}( I_{4\times 4}-i\tau^3-i\tau^2-i\tau^1)]$. From Sec.~\ref{generalities} we know that $\mathcal C_0: \Phi\rightarrow \pm O_T \Phi^\dagger$, where the overall phase depends on convention. Fixing the overall phase we get the last column in table \ref{table:triangular_monopole}.

\item{ Reflection $R_x$: \be \psi\xrightarrow{R}\frac{(\gamma_1-\sqrt{3}\gamma_2)}{2}\{i\sigma^y  W_c\tau^2\}\tau^2 \frac{\tau^3-\tau^1}{\sqrt{2}}\psi^*.
\ee
 This is really a $\mathcal{CR}$ symmetry. Since quantum spin hall mass $\overline \psi\sigma\psi$ stay invariant under $R$, we could invoke the reflection-protected topological phase argument in Sec ~\ref{dimred}. It is easy to check that $R^2=(-1)^F$, which means that $R: \Phi_{4/5/6}\to \Phi_{4/5/6}$  since this is not a nontrivial $R$ protected topological phase. To fix how $\Phi_{1/2/3}$ transform requires a bit of caution: first, they differ from $\Phi_{4/5/6}$ transformations by a minus sign from $\mathcal C_0$ (the part encoded in big bracket in the transformation), on top of that they are rotated by $\tau^2 \frac{\tau^3-\tau^1}{\sqrt{2}}$. Combining these steps, one gets full reflection transformation.}
\end{itemize}

 So at this point the only unfixed phase factors are those associated with $T_{1,2}$. We now calculate these phases using the Wannier center technique developed in Sec.~\ref{wanniercenter}.

\subsubsection{Monopole angular momenta from Wannier centers}

Below, we calculate the Wannier centers in the triangular lattice setting to deduce the monopole angular momenta for rotations around different centers. We use the  approach described in Sec ~\ref{wanniercenter} and consider three kinds of $3$-fold rotations around site $C_3^\cdot$, upward triangle center $C_3^\triangle$ and downward triangle center $C_3^\triangledown$, respectively (we omit $C_6$, since it changes the staggered flux pattern). Employing the dimensional reduction principles in Ref \onlinecite{song_2017} we first find the high symmetry points and calculate representations for the rotation groups. 

\begin{figure}[htbp]
\begin{center}
\captionsetup{justification=raggedright}
       \includegraphics[width=0.45\textwidth]{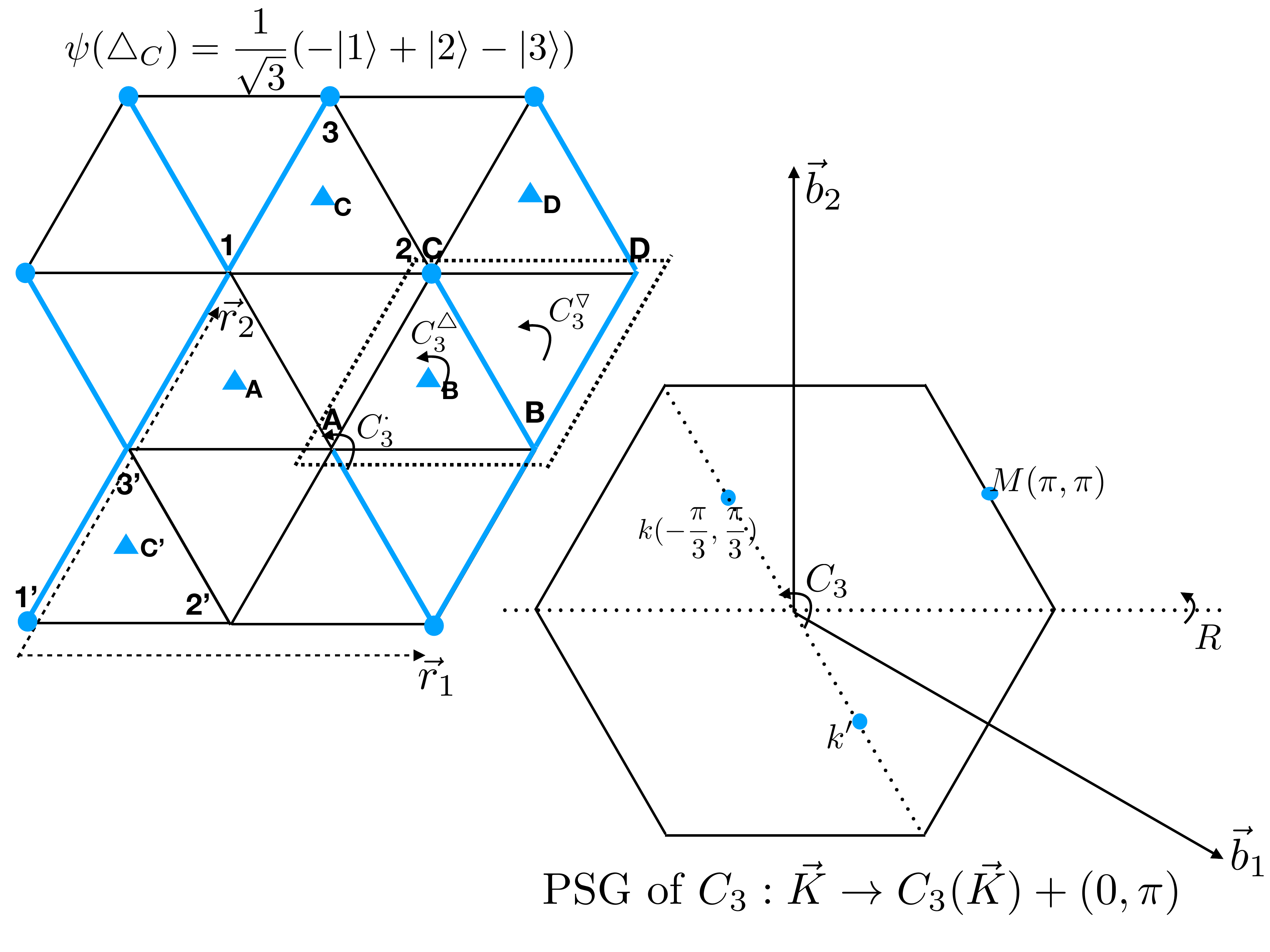}
 \caption{On the left is the 4-site unit cell used to calculate Wannier centers on the triangular lattice. Right: Brillouin zone of the 4-site unit cell Parton Hamiltonian. The $-1$ gauge transformation for $C_3$'s is labeled by a solid blue circle on site (those without the circle have trivial gauge transformation). This gauge transform has momentum $(0,\pi)$. Right panel is the rotation invariant Brillouin zone for the $4$-site unit cell. Marked are 3 rotation invariant momenta. }\label{fig:tri_psg}
 \end{center}
 \end{figure}

First one takes a $4$-site unit cell with a $C_3$ invariant Brilloun zone illustrated in figure \ref{fig:tri_psg}. Under the three types of $3$-fold rotations, the microscopic spinon fields transform as
\be
\phi_{(\vec r,i)}\rightarrow g[C_3(\vec r,i)] \phi_{C3(\vec r,i)}
\ee
where $\vec r$ labels Bravais lattice vector and $i$ labels the $A,B,C,D$ sublattices within a unit cell. The accompanying gauge transformation $g[\vec r,i] =g[0,i]exp (i \delta k\cdot (\vec r))$, i.e., the gauge transformation has momentum $\delta k=(0,\pi)$ shown in figure \ref{fig:tri_psg}. The momenta transform under $C_3$'s as 
\be 
C_3: \vec k\rightarrow C_3(\vec k) +(0,\pi)
\ee
which leads to three rotation invariant momenta, $M=(\pi,\pi)$ which is the $4$-fold degenerate Dirac point, and $k=(\frac{-\pi}{3},\frac{\pi}{3}),k'=-k$ (right panel of figure \ref{fig:tri_psg}). One could diagonalize the $C_3$ rotation matrix at these high symmetry points and the eigenvalues are listed in the last two columns of table \ref{table:tri_rep}.

To find the Wannier limit of the spinon band, we compare the band representation to those of the Wannier insulator centered on site, upward triangle and downward triangle, respectively. The representation of site-centered Wannier insulator is straightforward from the projective symmetry group since we are using wavefunction localized on site as our basis after all. For the other two types of Wannier limit, we first define the localized fermionic wave function basis. We stipulate that the wannier function localized at the center of certain plaquette to be a equal-amplitude superposition of the wave functions on site surrounded the plaquette. Then each Wannier center holds $n$ linearly independent wave functions where $n$ is the number of vertices on the boundary of the plaquette. \footnote{For the edge centered Wannier function on square and honeycomb lattice, the Wannier function is a superposition of wavefunctions on the endpoints of the edge.} In principle, one could use this new basis, diagonalize the $C_3$ matrix at high symmetry point to calculate the representation. Next we present a more direct and physical way to obtain the representation.

Consider site centered $C_3^v$ operation for upward triangle centered Wannier basis. As shown in figure \ref{fig:tri_psg}, we take the four upward triangles with the lower right site as reference point to form the unit cell marked by $\triangle_{A/B/C/D}$, it's obvious that under $C_3$ around site $A$ $\triangle_{A/B/D}$'s permutes to one another, since $(C_3^v)^3=1$, the matrix for these three triangles is always of the form (under appropriate phase choice of the superposition coefficients)
\be
\left(\begin{array}{ccc} 0&1&0\\ 0&0&1\\1&0&0\end{array}\right )
\ee
whose eigenvalue is $\{1,\omega,\omega^2\},(\omega\equiv exp(i\frac{2\pi}{3}))$, the $9$-dimensional subspace spanned by Wannier functions of $\triangle_{A/B/D}$ at high symmetry points therefore constitute a representation with eigenvalues $3\{1,\omega,\omega^2\}$, where $3$ means direct sum of the set of eigenvalues.  $\triangle_C$, on the other hand, goes to its equivalent under $C_3^\cdot$. Consider the Wannier function localized at $\triangle_C$:
\be
\psi(\triangle_C)=\frac{1}{\sqrt{3}} (-|1\rangle+|2\rangle-|3\rangle)
\ee
where the site labels are marked in figure \ref{fig:tri_psg}, under $C_3^\cdot$,
\be
\psi(\triangle_C)\rightarrow \psi(\triangle_{C'})=-\frac{1}{\sqrt{3}} (-|1'\rangle+|2'\rangle-|3'\rangle).
 \ee
 Since $C,C'$ differs by a lattice vector $\vec r_2$, the eigenvalue reads $-exp (i \vec k\cdot \vec r_2)$. Similarly for the other two $\triangle_C$ wave functions
 \be 
 \psi(\triangle_C)=\frac{1}{\sqrt{3}} (-|1\rangle+\omega^\eta|2\rangle-\omega^{2\eta}|3\rangle) \quad (\eta=1,2)
 \ee
 the eigenvalues are $-\omega^\eta exp (i \vec k\cdot \vec r_2)$.

 Since the Wannier insulator has to respect  rotation symmetry, each legitimate representation should involve all $4$ types of wannier centers $\triangle_{A/B/C/D}$, namely one should combine the above block diagonalized representation for $\triangle_{A/B/D}$ with some $\triangle_C$ wavefunction of certain angular momentum $L$, leading to the second column of table \ref{table:tri_rep}. We proceed in a similar fashion to produce the rest of the table. 
 
  \begin{table*}[ht]
  \renewcommand{\arraystretch}{1.4}

 \captionsetup{justification=raggedright}
\begin{center}
\begin{tabular}{|c|c|c|c|c|c|}
\hline
\hline
sym. & $\Gamma^\triangle_{k/M}$&$\Gamma^\triangledown_{k/M}$&$\Gamma^\circ_{k/M}$ &$\Gamma^{PSG}_k$({2-fold particle},{2-fold hole})&$\Gamma^{PSG}_M$(4-fold deg.)\\
\hline
$C_3^{v,\chi_c=0}$ & $-\omega^\eta p_1^*\oplus[1,\omega,\omega^2]$& $-\omega^\eta p_1\oplus[1,\omega,\omega^2]$&$1\oplus[1,\omega,\omega^2]$& $\Gamma^{particle}_k=[\omega^2,1],\Gamma^{hole}_k=[\omega,1]$& $[1,1,\omega,\omega^2]$\\
$C_3^{u,,\chi_c=1}$ & $\omega^\eta\oplus[1,\omega,\omega^2]$& $-\omega^\eta p_1^*\oplus[1,\omega,\omega^2]$&$-p_1\oplus[1,\omega,\omega^2]$& $\Gamma^{particle}_k=[\omega,1],\Gamma^{hole}_k=[\omega,\omega^2]$& $[1,1,\omega,\omega^2]$\\
$C_3^{d,,\chi_c=2}$ & $-\omega^\eta p_1\oplus[1,\omega,\omega^2]$& $\omega^\eta\oplus[1,\omega,\omega^2]$&$- p_2\oplus[1,\omega,\omega^2]$& $\Gamma^{particle}_k=[\omega^2,\omega],\Gamma^{hole}_k=[\omega^2,1]$& $[1,1,\omega,\omega^2]$\\
\hline
\multicolumn{2}{|c|} {$\Gamma^{particle}_k=2\{\omega^{\chi_c}[\omega^2,1]\}$}&\multicolumn{2}{c|}{$\Gamma^{particle}_M=[1,1,\omega,\omega^2]$}&\multicolumn{2}{c|}{$\Gamma^{particle}_{k,M} =\Gamma^\circ_{k,M}+\Gamma^\triangle_{k,M}-\Gamma^\triangledown_{k,M}$}\\
\multicolumn{2}{|c|} {$\Gamma^{hole}_k=2\{\omega^{\chi_c}[\omega,1]\}$}&\multicolumn{2}{c|}{$\Gamma^{hole}_M=[1,1,\omega,\omega^2]$}&\multicolumn{2}{c|}{$\Gamma^{hole}_{k,M} =\Gamma^\circ_{k,M}+\Gamma ^\triangledown_{k,M}-\Gamma^\triangle_{k,M}$}\\
\hline
\hline
\end{tabular}
\caption{The point space group representation of $C_3$ rotations (superscript v,u,d denotes rotation center as site, upward triangle and downward triangle center, respectively, $\chi_c$ is an index assigned to each rotation for notation convenience.) on triangular lattice at high symmetry points $k=(\frac{-\pi}{3},\frac{\pi}{3}),M=(\pi,\pi)$. $p_{1/2}=\exp(i \vec k\cdot r_{1/2})$ is the phase factor along $T_{1/2}$ translations at the high-symmetry momenta of interest. $\Gamma^\triangle,\Gamma^\triangledown,\Gamma^\circ$ denotes representation of Wannier functions centered at upward,downward triangle and on site, respectively. $\omega=e^{i\frac{2\pi}{3}}$ and integer $\eta$ is related to orbital angular momentum of Wannier function and can differ from each to each ($\eta=0$ for site-center wannier function by our Hilbert space choice).The quantum spin hall mass will open a gap at $M$ point and the filled band for spin up and down are related by time reversal. Hence the filled band at Dirac point has eigenvalue $\{1,1,\omega,\omega^2\}$. \label{table:tri_rep}}

\end{center}
\end{table*}
 
 Compare the representation of spinon bands with those of the $3$ Wannier insulators, we identify a unique decompositon of the spinon bands into Wannier centers (lower left of Fig \ref{fig:honeycomb2}):
 \be
\textrm{triangular QSH occupied band =site+}\triangle-\triangledown
 \ee
 where again the minus sign denotes formal difference in light of fragile topology.  Taking into account the background gauge charge $-1$ per site, the spin triplet monopoles see no gauge charge rotating around site and $\pm 1$ gauge charge rotating around upward/downward triangle center, respectively. Since $T_2=(C_3^d)^{-1}C_3^u$ (d,u denotes rotations around downward,upward triangle centers, respectively), monopoles $\mathcal S_i$'s transform with a phase $-\frac{-2\pi}{3}+\frac{2\pi}{3}=\frac{4\pi}{3}$, and similarly they get a phase of $\omega$ under $T_1$. To sum up, we get the transformation of monopoles as tabulated in Table~\ref{table:triangular_monopole}. 
The minimal symmetry-allowed monopole is a three-fold monopole as discussed in Ref.~\cite{shortpaper}.
  
Note that in principle, it is not sufficient to simply match the eigenvalues because of the complicated PSG structure. A full calculation should compare the full representations of lattice PSG (rotation and translation) at high symmetry points in the Brillouin zone, which are generically non-Abelian. We do not perform such calculation here since the much simpler calculation is already sufficiently  constraining to essentially fix the Wannier centers.

\subsection{Kagom\'e lattice}
\label{kagome_calculation}

\begin{table}
\captionsetup{justification=raggedright}
\begin{center}

\begin{tabular}{|c|c|c|c|c|c|}
\hline
& $T_1$ & $T_2$ & $R_y$ & $ C_6$ & $\mathcal T$ \\
\hline
$\Phi_1^\dagger$ & $-\Phi_1^\dagger$ &$- \Phi_1^\dagger$ & $-\Phi_3$ & $e^{i\frac{2\pi}{3}}\Phi_2^\dagger$ & $\Phi_1$\\
$\Phi_2^\dagger$ & $\Phi_2^\dagger$ &$-\Phi_2^\dagger$ & $\Phi_2 $ & $-e^{i\frac{2\pi}{3}}\Phi_3^\dagger$ & $\Phi_2 $\\
$\Phi_3^\dagger$ & $-\Phi_3^\dagger$& $\Phi_3^\dagger$&$-\Phi_1 $ & $-e^{i\frac{2\pi}{3}}\Phi_1^\dagger$ & $\Phi_3 $\\
$\Phi_{4/5/6}^\dagger $ & $\Phi_{4/5/6}^\dagger $& $\Phi_{4/5/6}^\dagger $& $-\Phi_{4/5/6} $ &$e^{i\frac{2\pi}{3}}\Phi_{4/5/6}^\dagger  $ & $-\Phi_{4/5/6} $\\
\hline
\end{tabular}
\caption{The transformation of monopoles on kagom\'e lattice. $R_y,C_6$ denotes reflection perpendicular to vertical direction and six-fold rotation around center of hexagon in fig \ref{fig:honeycomb2}.  It's impossible to incorporate the $6$-fold rotation of monopoles to a vector representation of $SO(6)$ owing to the nontrivial Berry phase, which is in line with magnetic pattern on kagom\'e lattices. 
 \label{table:kagome_monopole}}
\end{center}
\end{table}

On kagom\'e lattice, 
similar to triangular case, Ref.~\onlinecite{hermele_2008} calculated the kagom\'e DSL with staggered flux mean-field ansatz, with three gamma matrices as $\gamma_\nu=(\mu^3,\mu^2,-\mu^1)$, and we have for the PSG of Dirac fermions as
\begin{align}
T_1: \psi\rightarrow (i\tau^2)\psi\quad T_2: \psi\rightarrow (i\tau^3)\psi\quad R_y: \psi\rightarrow (i\mu^1)e^{\frac{i\pi}{2}\tau_{ry}} \psi\nonumber\\
C_6: \psi\rightarrow e^{\frac{i\pi}{3}\mu^3} e^{\frac{2\pi i}{3}\tau_R}\psi\quad \mathcal T:\psi\rightarrow (i\sigma^2)(i\mu^2)(-i\tau^2)\psi.
\end{align}
where
\begin{align}
\tau_{ry}=\frac{-1}{\sqrt{2}} (\tau^1+\tau^3)\quad \tau_R=\frac{1}{\sqrt{3}}(\tau^1+\tau^2-\tau^3).
\end{align}

The PSG again fixes the monopole transformations, up to overall phase factors to be determined, as listed in Table~\ref{table:kagome_monopole}. The overall phase for time-reversal is fixed through the argument in Sec.~\ref{generalities}. The overall phase for $R_y$ is convention-dependent and we fix it as in Table~\ref{table:kagome_monopole}. The fact that $\Phi_{4,5,6}$ are invariant (up to a phase) under $C_6$ requires their momenta to be zero, which in turn fixes the overall phases associated with $T_{1,2}$ as in Table~\ref{table:kagome_monopole}. The only undetermined phase is that in $C_6$, which we fix below. The calculation is essentially the same as in the triangular case, so we will be brief here.

One calculates symmetry representations of $3$-fold rotation around upward/downward triangle centers, and $6$-fold rotation around hexagon centers, listed in table \ref{table:kagome_rep}. The spinon band is represented as (lower right of Fig \ref{fig:honeycomb2})
 \be
\textrm{kagome QSH occupied band= 3 site} -\triangle -\triangledown- 4\hexagon
 \ee
 where $\triangle,\triangledown,\hexagon$ denotes Wannier insulators localized on upward/downward triangles, and hexagons. Note the numeral factor only denotes the numbers of occupied  particles localized at Wannier center, they may have different Wannier wave functions. This means the $C_6$ rotation sees a $\frac{-4\pi}{3}$ Berry phase, and translation begets no Berry phase since translation is composed of $C_3^\triangle (C_3^\triangledown)^{-1}$ and the two Berry phase cancels. These are the results listed in Table~\ref{table:kagome_monopole}.
The most relevant (in the RG sense) symmetry-allowed monopole is a two-fold monopole as discussed in Ref.~\cite{shortpaper}. 

\begin{table*}[]
\renewcommand{\arraystretch}{1.4}
\captionsetup{justification=raggedright}
\begin{center}
\begin{tabular}{|c|p{18mm}|p{18mm}|p{30mm}|p{18mm}|p{40mm}|p{40mm}|}
\hline
sym. & $\Gamma^\triangle_{k/Q}$&$\Gamma^\triangledown_{k/Q}$&$\Gamma^{\hexagon}_{k/Q}$ &$\Gamma^\circ_{k/Q}$ &$\Gamma^{PSG}_k$({4-fold particle,2-fold particle,4-fold hole,2-fold hole})&$\Gamma^{PSG}_Q${4-fold particle,4-fold(Dirac fermion), 4-fold hole}\\
\hline
$C_3^{u}$ & $\omega^\eta\oplus[1,\omega,\omega^2]$ &$\omega^\eta p_1\oplus[1,\omega,\omega^2]$& $\omega^\eta p_1^*\oplus[1,\omega,\omega^2]$ & $4[ 1,\omega,\omega^2]$ & $[(\omega,\omega^2),(1,\omega)]\oplus[1,\omega^2]\oplus[(\omega,\omega^2),(1,\omega)]\oplus[1,\omega^2]$ & $[\omega,\omega,\omega^2,\omega^2]\oplus [1,1,\omega,\omega^2] \oplus [1,1,\omega,\omega^2]$\\
$C_3^{d}$ & $\omega^\eta p_1^*\oplus[1,\omega,\omega^2]$ &$\omega^\eta \oplus[1,\omega,\omega^2]$& $\omega^\eta p_1\oplus[1,\omega,\omega^2]$ & $4[ 1,\omega,\omega^2]$ & $[(1,\omega^2),(\omega^2,\omega)]\oplus[1,\omega]\oplus[(1,\omega^2),(\omega^2,\omega)]\oplus[1,\omega]$ & $[\omega,\omega,\omega^2,\omega^2]\oplus [1,1,\omega,\omega^2] \oplus [1,1,\omega,\omega^2]$\\
$C_3^{h}$ & $\omega^\eta p_1\oplus[1,\omega,\omega^2]$ &$\omega^\eta p_1^*\oplus[1,\omega,\omega^2]$& $\omega^\eta \oplus[1,\omega,\omega^2]$ & $4[ 1,\omega,\omega^2]$ & $[(\omega,1),(1,\omega^2)]\oplus[\omega,\omega^2]\oplus[(\omega,1),(1,\omega^2)]\oplus[\omega,\omega^2]$ & $[\omega,\omega,\omega^2,\omega^2]\oplus [1,1,\omega,\omega^2] \oplus [1,1,\omega,\omega^2]$\\
\hline
$C_6$ &\multicolumn{2}{c|} {$[\Omega^{2\eta+1} ,-\Omega^{2\eta+1} ]\oplus [\textrm{Sextet}]$} &$\Omega^{2\eta+1} \oplus(-1)^{\eta}[e^{i\frac{\pi}{6}},e^{i\frac{5\pi}{6}},-i]$ &$2[\textrm{Sextet}]$ &\multicolumn{2}{c|} {$\Gamma^{PSG}_Q: [ \Omega,\Omega^5,\Omega^{-5},\Omega^{*}]\oplus [\pm i,\Omega,\Omega^{*}]\oplus [ \pm i,\Omega^5,\Omega^{-5}] $}\\
\hline
\multicolumn{7}{|c|} {$\Gamma^{particle}=3\Gamma^\circ_{L=0}-\Gamma^\triangle_{L=0}-\Gamma^\triangledown_{L=0}-\Gamma^{\hexagon}_{L=0}-\Gamma^{\hexagon}_{L=3}-\Gamma^{\hexagon}_{L=4}-\Gamma^{\hexagon}_{L=5}$}\\
\multicolumn{7}{|c|} {$\Gamma^{hole}=-\Gamma^\circ_{L=0}+\Gamma^\triangle_{L=0}+\Gamma^\triangledown_{L=0}+\Gamma^{\hexagon}_{L=0}+\Gamma^{\hexagon}_{L=3}+\Gamma^{\hexagon}_{L=4}+\Gamma^{\hexagon}_{L=5}$}\\
\hline
\end{tabular}
\end{center}
\caption{The space group representation of rotations on kagom\'e lattice at high symmetry points $k=(\frac{2\pi}{3},\frac{\pi}{3}),Q=(0,\pi)$. 
Superscript u,d,h denotes rotation around upward, downward triangle and hexagon centers, respectively.
We have $\Omega=e^{i\frac{\pi}{6}}$, $\omega=e^{i2\pi/3}$, $[\textrm{Sextet}]=i[e^{\pm i\frac{\pi}{6}},\pm i,e^{\pm i\frac{5\pi}{6}}]$, and $\eta$ is an integer and can vary from each to each. $p_{1}=e^{i \vec k\cdot r_{1}}$ is the phase factor under $T_1$ translation at the high-symmetry momenta of interest. 
$\Gamma^\triangle,\Gamma^\triangledown,\Gamma^\circ, \Gamma^{\hexagon}$ denote representation of Wannier functions centered at upward,downward triangle, sites and hexagon, respectively. 
At $k$ point, there are two sets of $4-$fold degenerate states, which can be simultaneously block diagonalized into two $2\times 2$ matrices for all three $C_3$ rotations and we list the pairwise eigenvalues in brackets. The multiplicity of reps. is denoted as simply a number in front of the set of eigenvalues. }
\label{table:kagome_rep}
\end{table*}

\section{Anomalies and Lieb-Schultz-Mattis}
\label{anomalyLSM}

In $CP^1$ (slave boson) representation of spin-$1/2$ systems on square lattice, the monopole quantum numbers were fixed by Lieb-Shultz-Mattis (LSM) constraints\cite{MetlitskiThorngren}. Essentially, the LSM theorem requires the low energy effective field theory to have certain symmetry anomalies, which can be matched only if the $CP^1$ monopole carries the right quantum number, e.g. $\pm1$ angular momentum under $C_4$. It is then natural to ask to what extent are the monopole quantum numbers in Dirac spin liquids determined by LSM-anomaly constraints. In this section we show that monopole quantum numbers associated with $\mathbb{Z}_2$ symmetries (such as inversion) are indeed determined by LSM-anomaly constraints in DSL, while those associated with $\mathbb{Z}_3$ symmetries (such as $C_3$) are not.

Let us start from the QED$_3$ theory, and try to gauge the $SO(3)_{spin}\times SO(3)_{valley}\times U(1)_{top}$ symmetry\footnote{Notice that this symmetry is simpler than $SO(6)\times U(1)/\mathbb{Z}_2$ since $SO(3)\times SO(3)$ has no center.}, as we will be interested in those microscopic symmetries that can be embedded into this group. The anomaly associated with these symmetries can be calculated. One way to interpret the anomaly is to imagine a $(3+1)d$ SPT state that hosts the QED$_3$ theory on its boundary, and the bulk SPT is characterized by a response theory
\be
\label{Anomaly}
S_{bulk}=i\pi\int_{X_4}\left[ w_2^s\cup w_2^v+\left(w_2^s+w_2^v+\frac{dA_{top}}{2\pi} \right)\cup \frac{dA_{top}}{2\pi}  \right],
\ee
where $w_2^{s,v}$ are the second Stiefel-Whitney classes of the $SO(3)_{spin/valley}$ bundles, respectively, and $A_{top}$ is the $U(1)$ gauge field that couples to the $U(1)_{top}$ charge. We outline the derivation of the above expression (which is similar to that in Ref.~\cite{wang_2017}) in Appendix~\ref{SSUanomaly}. The first term is essentially a descendent of the parity anomaly of Dirac fermions, and the terms involving $A_{top}$ simply represents the fact that an $A_{top}$ monopole -- the spinon $\psi$ in the original QED$_3$ -- carries half-spin under $SO(3)_{spin/valley}$ and is a fermion. 

Our remaining task is simply to embed the lattice symmetries into the $SO(3)_{valley}\times U(1)_{top}$ group and see if the correct anomaly from LSM is reproduced. The PSG determines the $SO(3)_{valley}$ part of the lattice symmetries, which in turn determines $w_2^v$. The only unknown is the relation between the lattice symmetries and $A_{top}$, determined by the $U(1)_{top}$ Berry phase in the symmetry realizations.

Let us first consider $\mathbb{Z}_2$ inversion symmetries. On triangular lattice this is not interesting since it involves charge conjugation. We shall consider the other three lattices in detail.

On square lattice the site-centered inversion involves a nontrivial $SO(3)_{valley}$ rotation $\psi\to i \tau^2\psi$. So if we gauge the inversion symmetry (call the $\mathbb{Z}_2$ connection $\gamma$), the $\pi w_2^s w_2^v$ term in Eq.~\eqref{Anomaly} becomes $\pi w_2^s\gamma^2$(short hand for $\gamma\cup\gamma$), which is exactly the anomaly imposed by LSM (simply reflecting the fact that there is a spin-$1/2$ Hilbert space at the inversion center on lattice). Therefore the other terms in Eq.~\eqref{Anomaly} should not contribute further anomalies. This means that $dA_{top}=0$, i.e. there is no additional $U(1)_{top}$ phase factor associated with inversion. This is indeed what we have in Table~\ref{table:square_monopole}, where inversion only implements the $SO(3)_{valley}$ rotation by $\Phi_{1,3}\to-\Phi_{1,3}$.

On honeycomb lattice the $\pi w_2^s w_2^v$ term likewise gives a $\pi w_2^s\gamma^2$ anomaly where $\gamma$ is again the inversion gauge field. However, since on the lattice there is no spin at the inversion center, there should be no actual anomaly. This means that the other terms in Eq.~\eqref{Anomaly} should contribute another $\pi w_2^s\gamma^2$ term to the anomaly, which can be done by having $dA_{top}=2\pi\gamma^2$. This means that under inversion the monopole should pick up an additional $(-1)$ phase, exactly in accordance with Table~\ref{table:honeycomb_monopole}.

On kagome lattice, the hexagon-centered inversion does not involve any nontrivial $SO(3)_{valley}$ rotation for the Dirac fermions, therefore the $\pi w_2^s w_2^v$ term does not contribute an anomaly for inversion. The LSM constraint also requires no anomaly since on the lattice there is no spin at the inversion center. Therefore the terms involving $A_{top}$ in Eq.~\eqref{Anomaly} should not give rise to any anomaly either. This means that under inversion the monopoles stay invariant, again in accordance with Table~\ref{table:kagome_monopole}.

We now consider $\mathbb{Z}_3$ (or any $\mathbb{Z}_{2k+1}$) symmetries like the $C_3$ rotations. In this case the anomalies become trivial no matter how we embed the symmetry to $SO(3)_{valley}\times U(1)_{top}$. This can be seen most easily by writing the anomalies involving $SO(3)_{spin}$ as $\pi w_2^s w_2^{SO(3)_{valley}\times U(1)_{top}}$ (recall that $w_2^{SO(2)}=dA/2\pi$ (mod $2$)), and using the fact that $w_2=0$ for a $\mathbb{Z}_3$ bundle.  Therefore the anomaly-based argument does not say anything about monopole quantum numbers for these symmetries.

As a final example, let us consider translation symmetries $T_{1,2}$ on triangular lattice. These symmetries act as $\mathbb{Z}_2\times \mathbb{Z}_2$ on Dirac fermions, but could also involve a $\mathbb{Z}_3$ subgroup of $U(1)_{top}$ (one can show translation involves Berry phase $2n\pi/3(n\in\mathbb Z)$ from algebraic relations of space group\cite{shortpaper}). The $\mathbb{Z}_2\times \mathbb{Z}_2$ part gives an anomaly $\pi w_2^{s}xy$ where $x,y$ are the $\mathbb{Z}_2$ forms associated with $T_{1,2}$, and this is exactly the LSM anomaly since we have one spin-$1/2$ per unit cell. The $\mathbb{Z}_3$ part, however, will not further modify the anomaly, so the two in-equivalent choices of Berry phase ($e^{i2\pi/3}$ or trivial) are both allowed by LSM.

\section{Three dimensions: Monopole PSG from charge centers}
\label{3Dchargecenter}

The charge (Wannier) center approach can also be generalized to three dimensions. Consider a $3D$ $U(1)$ quantum spin liquid with gapped spinons (charge) and magnetic monopoles. Recall that in $3D$ a monopole is a point excitation, and the monopole creation operator is a nonlocal operator. Such spin liquids have been extensively discusses in the context of quantum spin ice materials\cite{savary_2017}. The non-local nature of monopoles in $3D$ implies that unlike the 2D case, they can transform projectively under physical symmetries like the spinons. The relevant  question then is how the spinon band topology (or SPT-ness) affects the monopole projective symmetry quantum number (or simply monopole PSG).

We now show that the monopole PSG associated with lattice rotation symmetries are determined by the effective gauge charges sitting at the rotation centers, in cases when the rotation symmetries do not involve charge conjugation (namely when the monopole flux is invariant under the lattice rotations). To see this, first consider a single charge, or any odd-integer charge, in space with the full $SO(3)$ rotation symmetry around the charge. By examining the Aharonov-Bohm phase for a monopole moving around the charge, one concludes that the monopole carries half-integer angular momentum under the space $SO(3)$ rotation\footnote{One choice of vector potential for a $2\pi$ Dirac monopole reads $\vec A =\frac{(1-\cos(\theta))}{2r\sin(\theta)}\hat\varphi$ in polar coordinates parametrized by $(r,\theta,\varphi)$, on the unit sphere identical to Berry connection for spin-$1/2$.} i.e. the monopole transforms projectively under $SO(3)$. Now on a lattice the $SO(3)$ is broken down to a discrete subgroup, but as long as the remaining rotation group $G$ admits a projective representation $\omega_2\in H^2(G,U(1))$ that is a descendent of the spin-$1/2$ representation when $G$ is embedded into $SO(3)$, the monopole will transform projectively under $G$ according to $\omega_2$.

The simplest examples is the dihedral group $D_2=\mathbb{Z}_2\times \mathbb{Z}_2$, corresponding to $\pi$ rotations about three orthogonal axes. If an odd number of gauge charges effectively sit at the rotation center, the monopole will transform under $D_2$ such that different $\mathbb{Z}_2$ rotations anti-commute.

\section{Discussion and Future Directions}
\label{Discussion}
We have demonstrated a precise mapping between Landau order parameters and the symmetry protected band topology of fermions. The link is established by studying the properties of magnetic monopoles in a Dirac spin liquid. Knowledge of the spinon band topology allows us to analytically calculate monopole symmetry quantum numbers. This in turn allows us to identify  the set of order parameters that are enhanced in the vicinity of a Dirac spin liquid.  Thus, results involving  gapped and noninteracting  fermions, which is an  analytically well controlled limit, are used to extract key information about a strongly interacting and gapless system, the Dirac spin liquid. We also showed that on bipartite lattices there always exists a symmetry-trivial monopole due to the existence of a parent SU(2) gauge theory. In a separate publication we have discussed the physical consequence of the monopole properties as well as signatures of a Dirac spin liquid which can be accessed in numerics and in scattering experiments on candidate materials\cite{shortpaper}. 

Even though all our calculations of the Wannier centers were done in free fermion framework, notions like the monopole quantum numbers are certainly well-defined beyond free fermions. In fact,  the connection between Wannier centers and monopole quantum numbers gives intrinsically interacting definitions of notions like Wannier centers and fragile topology (in accordance with the recent work\cite{ElsePoWatanabe,shiftinsulator}).

The monopole angular momenta arising from charge (Wannier) centers are consistent with another simple fact in lattice gauge theory: in the strong-coupling limit of a lattice gauge theory (which is fully confined), all the gauge charges simply sit on the defining sites of lattice and do not fluctuate. Therefore if the gapped charge fields have their effective charge centers (Wannier centers) off the sites (say at plaquette centers), the strong-coupling limit must destroy such a state and pull the charges back to the lattice sites, implying that the lattice symmetries (that prevent charges from moving away from nontrivial centers) must be broken along the way. 
    
    One can use this as an intuitive way to understand the following well-known statement: the $SU(2)$ Yang-Mills theory in $(3+1)$ dimensions with $\theta=\pi$ cannot confine to a trivial phase without breaking time-reversal symmetry (for a recent exposure see Ref.~\onlinecite{GKKS}). When put on the lattice, $\theta=\pi$ can be generated through introducing fundamental fermions and putting the fermions into a topological superconductor state (protected by $SU(2)\times\mathcal{T}$). By definition the topological superconductor cannot be deformed (without closing gap or breaking symmetries) to a trivial product state with local $SU(2)$ singlets on lattice sites -- but this is exactly what the strong-coupling limit demands. Therefore if the theory indeed flow to infinite coupling, without closing the fermion gap, then time-reversal symmetry must be broken in the IR. 
    
We end with a discussions of open issues:
\begin{itemize}
    
    
    \item 
    It is also straightforward to apply the charge center method to $\mathbb{Z}_2$ spin liquids (or other discrete gauge theories) in $(2+1)d$. For example, if a nontrivial $\mathbb{Z}_2$ charge (effectively) sits at the center of a plaquette, then the gauge flux (vison) will transform projectively under the lattice rotation around the plaquette center. This is similar to the situation for $3D$ $U(1)$ spin liquids discussed in Sec.~\ref{3Dchargecenter}.

    \item One natural question is: given a microscopic Hilbert space (say, one spin-$1/2$ per site on triangular lattice), is it possible to realize a different $U(1)$ Dirac spin liquid, with the same field content and symmetry realizations (PSG), but different monopole quantum number from the one discussed in this work? For example, is there a spinon mean field ansats that gives identical Dirac dispersion and PSG, but with monopoles at different momenta, say $T_{1,2}: \Phi_{4/5/6}\to\Phi_{4/5/6}$ (as opposed to that in Table~\ref{table:triangular_monopole})? Unlike the bosonic spinon theory for deconfined criticality, some of the monopole quantum numbers in our examples are not linked to Lieb-Schultz-Mattis (LSM) constraints\cite{MetlitskiThorngren}, as discussed in Sec.~\ref{anomalyLSM}. So an alternative $U(1)$ Dirac spin liquid (with the same PSG but different monopole quantum numbers) seems allowed on general ground -- finding a concrete  example is left for future work.
    
    \item In Sec.~\ref{wanniercenter} and \ref{calculation} we calculated the angular momenta of monopoles using charge centers, and then obtained their momenta by composing different rotations. In the absence of lattice rotation symmetries the angular momenta are not defined, but the momentum of a monopole is still well defined (although may not be quantized). It is then natural to ask what determines monopole momentum when lattice rotation symmetries are not considered. Recall that with rotation symmetries, the momentum is given by the difference between angular momenta around different rotation centers, which is in turn given by the difference between gauge charges sitting at different rotations centers. The latter appears to be nothing but the dipole moment in the unit cell, which in the more familiar language is just the polarization density. In fact by comparing Fig.~\ref{fig:honeycomb2} and the monopole momenta for each spinon insulator, we can infer that monopole momentum $\vec{k}_{\mathcal{M}}$ is given by the polarization density $\vec{P}$ (dipole density per unit cell) through $\vec{k}_{\mathcal{M}}=2\pi\hat{z}\times\vec{P}$. In a forthcoming work we will develop the connection between polarization and monopole momentum in detail without assuming lattice rotation symmetries\cite{toappear}.
    
\end{itemize}

{\it Note added:} In a different parallel work \cite{shiftinsulator},  monopole quantum numbers were used as a probe of free fermion band topology for a specific class of models. Here however,  monopoles are dynamical objects, and the underlying models are of strongly interacting quantum magnets.  
\section*{Acknowledgements}

  We gratefully acknowledge helpful discussions with Chao-Ming Jian, Eslam Khalaf, Shang Liu, Jacob McNamara, Max Metlitski, Adrian H. C. Po, Ying Ran, Subir Sachdev, Cenke Xu, Yi-Zhuang You and Liujun Zou. X-Y~S acknowledges hospitality of Kavli Institute of Theoretical Physics (NSF PHY-1748958). YCH was supported by the Gordon and Betty Moore Foundation under the EPiQS initiative, GBMF4306, at Harvard University. A.V. was supported by a Simons Investigator award. CW was supported by the Harvard Society of Fellows. Research at Perimeter Institute (YCH and CW) is supported by the Government of Canada through the Department of Innovation, Science and Economic Development Canada and by the Province of Ontario through the Ministry of Research, Innovation and Science.

\begin{widetext}
\appendix

\section{Mean field, Projective symmetry group and fermion bilinear transformations}
\label{bilinears}
\subsection{Square}
We adopt a mean-field where $t_{ij}=(-1)^y$ for horizontal links and $t_{ij}=1$ for vertical links on square lattice which creates $\pi$ flux on every plaquette. This choice enlarges the unit cell to contain two sites (sublattice $A,B$) with a vertical link. There're two gapless points in the reduced Brillouin zone at $\bf Q=(\pi/2,\pi), \bf Q'=-\bf q$.

 The projective symmetry group reads (applicable also for staggered flux state)
 \begin{align}
 \label{square_psg}
T_1&: \psi\rightarrow i\mu^3\sigma^2\tau^3\psi^*\quad T_2: \psi\rightarrow i\mu^3\sigma^2\tau^1\psi^*\nonumber\\
R_x&: \psi\rightarrow \tau^3\mu^3 \psi\quad C_4: \psi\rightarrow \frac{1}{\sqrt{2}}\mu^3\sigma^2 (I-i\tau^2)e^{i\frac{\pi}{4}\mu^1} \psi^*\nonumber\\
\mathcal T&: \psi\rightarrow \tau^2 \mu^1 \psi^*\\
\mathcal C&: \psi\rightarrow i\mu^3\sigma^2\psi^*
\end{align}
where $\mathcal C$ denotes charge conjugation that reverses the flux $\phi\rightarrow -\phi$.

\subsection{Honeycomb}

On honeycomb lattice, with mean-field ansatz of uniform fermion hopping, one could similarly work out the PSG and the constraints on monopole quantum numbers. The Dirac points stay at momenta $\mathbf Q=(\frac{2\pi}{3},\frac{2\pi}{3}),\mathbf Q'=-\mathbf Q$.

  The physical symmetries act as

\begin{align}
\label{eqn:honey_psg}
T_{1/2}&: \psi\rightarrow e^{-i\frac{2\pi}{3}\tau^3} \psi\nonumber\\
C_6&: \psi\rightarrow -i e^{-i\frac{\pi}{6}\mu^3} e^{i (\cos{\frac{\pi}{3}}\tau^1+\sin{\frac{\pi}{3}} \tau^2)\frac{\pi}{2}} \psi \quad
R: \psi\rightarrow -\mu^2\tau^2 \psi\nonumber\\
\mathcal T&: \psi\rightarrow -i\sigma^2\mu^2\tau^2 \psi \quad
\text{charge conjugation } C: \psi\rightarrow \mu^1 \psi^*
\end{align}
where $T_{1/2}$ is the translation along two basis vectors with $2\pi/3$ angle between them,$C_6$ is $\pi/3$ rotation around a center of a honeycomb plaquette, and $R$ denotes reflection along the direction of the unit cell.

From the transformation of Dirac fermions, one gets transformation of fermion masses as table \ref{table:bilinears}.

\subsection{Triangular lattice}

There's a ``staggered $\pi$ flux" configuration of $t_{ij}$ on the triangular lattice. We choose a particular gauge of $t_{ij}$ to realize this mean field as in Fig \ref{fig:tri}. Under appropriate basis the low-energy Hamiltonian reads as the standard form with $4$ gapless Dirac fermions.

\begin{figure}[htbp] 
 \begin{center}
 \captionsetup{justification=raggedright}
 \adjustbox{trim={.0\width} {.13\height} {.15\width} {.16\height},clip}
 { \includegraphics[width=0.4\textwidth]{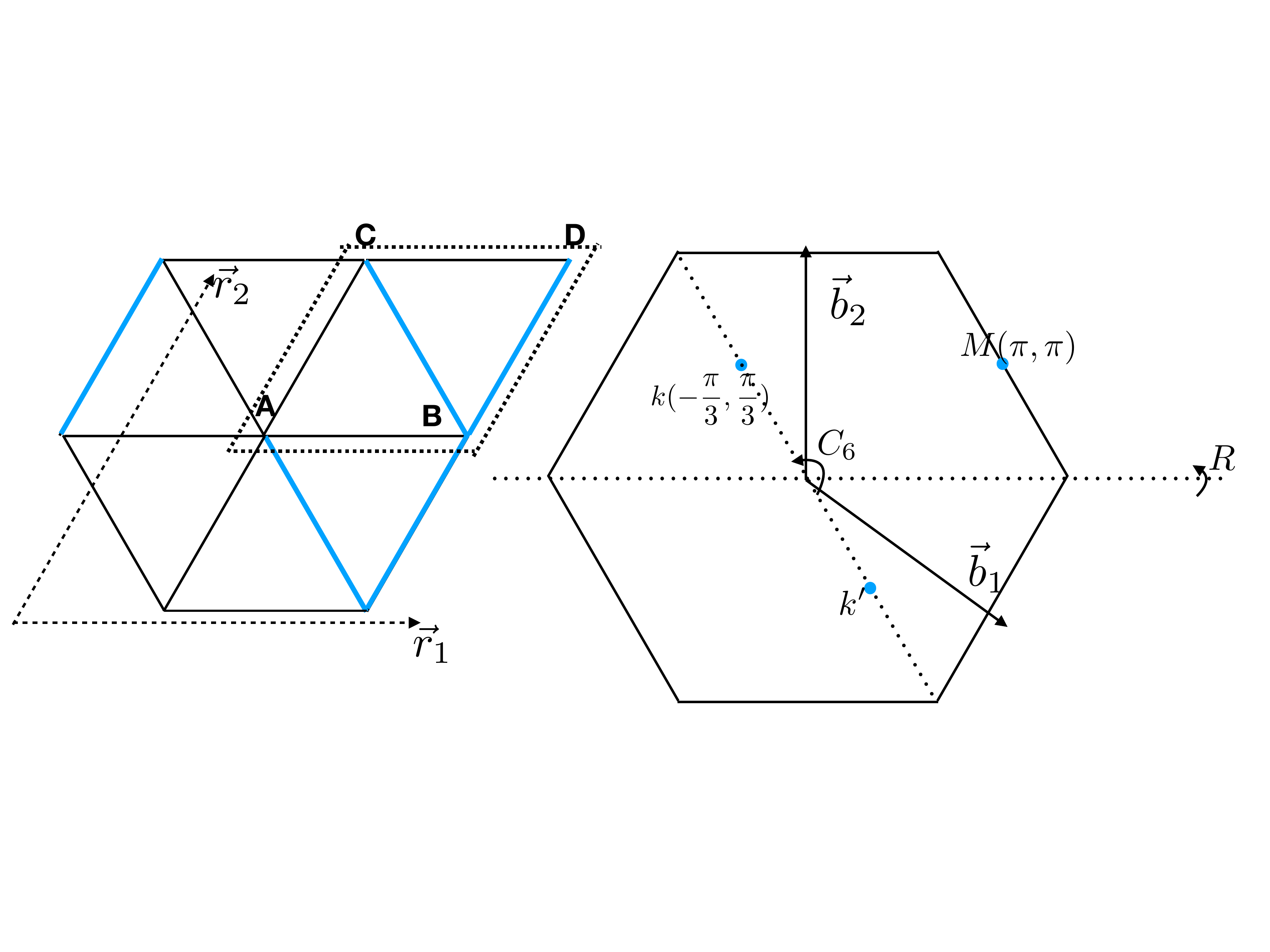}}
    \includegraphics[width=0.4\textwidth]{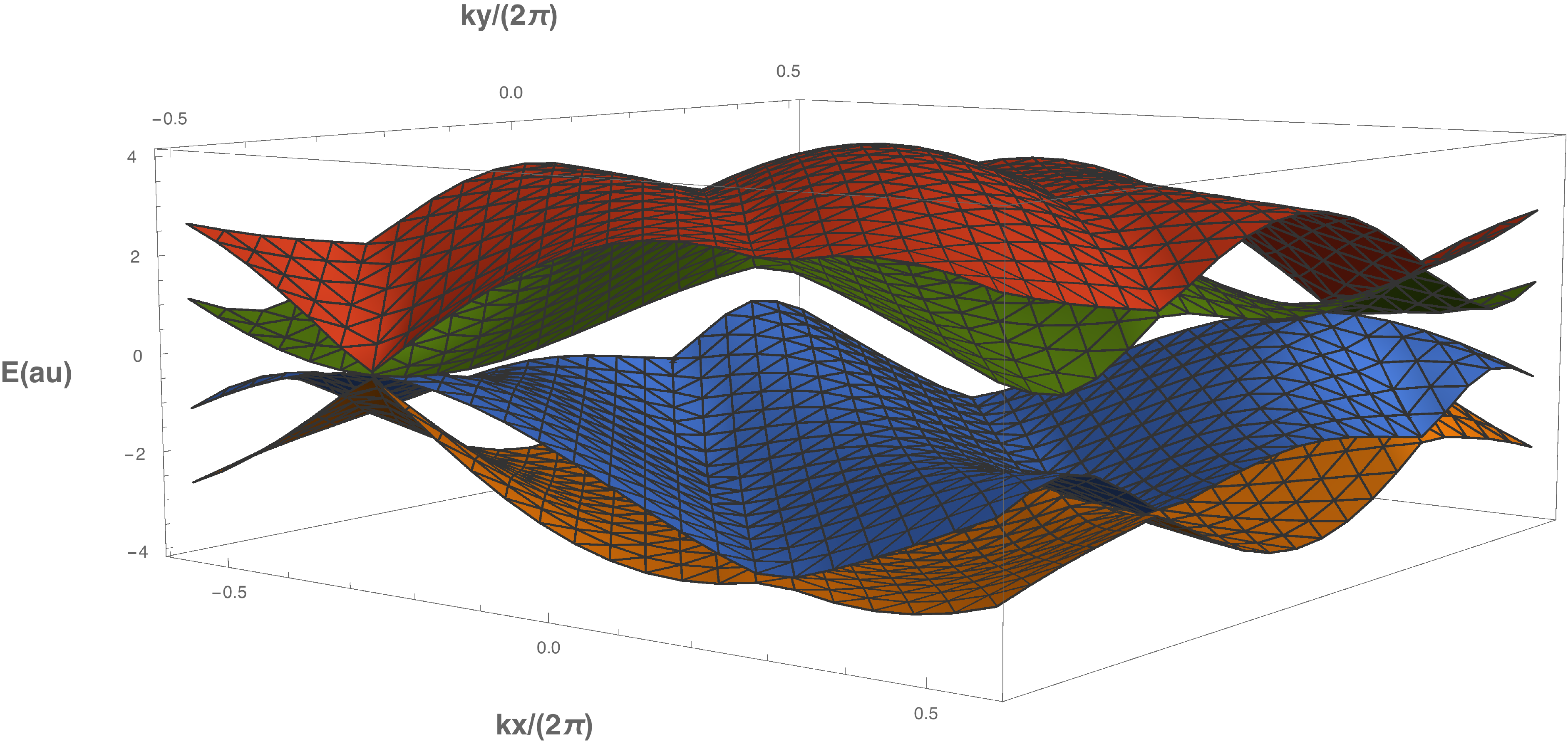}
 \caption{Left panel:The gauge choise (blue/black bonds denote negative/positve hopping strength with equal amplitudes.) and the Brillouin zone with $M$ point the gapless point and $k,k'$ stay invariant under PSG of $3$-fold rotations. Right panel: The energy spectrum of staggered flux spinon model on triangular lattice in reduced Brillouin zone with $4$ site unit cell. The spectrum is gapless at momentum $(\pi,\pi)$.
  }    \label{fig:tri}
 \end{center}
 \end{figure}

For later purposes, the charge conjugate operation is given here as
\be \psi\rightarrow W_c\psi^*=\frac{1}{\sqrt{3}} (-i I_{4\times4}-\mu^3+\mu^1) \psi^*\ee

The PSG for all the symmetry operations are the following:
\begin{eqnarray}
\psi&\xrightarrow{T_2}&i\tau^3\psi\quad\quad
\psi\xrightarrow{T_1}-i\tau^2\psi\quad\quad
\psi\xrightarrow{\mathcal T}i\sigma^2 \mu^2\tau^2\mathcal K\psi(-k)\\
\psi(k_1,k_2)&\xrightarrow{C_6}&i\sigma^2  W_{C_6} \psi^\dagger(-\frac{k_2}{2},2k_1-k_2)\quad\quad
\psi(k_1,k_2)\xrightarrow{R}i\sigma^y  W_{R} \psi^\dagger(k_1-\frac{k_2}{2},-k_2)
\end{eqnarray}
where 
\begin{eqnarray}
W_{C_6}=e^{-i\gamma^3 \frac{\pi}{6}} W_c exp[i\frac{\pi}{3}\tau^C]\quad \tau^C=\frac{1}{\sqrt{3}}(\tau^1+\tau^2+\tau^3)\nonumber \\
W_R=\frac{(\gamma^1-\sqrt{3}\gamma^2)}{2} W_c \frac{\tau^3-\tau^1}{\sqrt{2}}
\end{eqnarray}

\subsection{Kagom\'e}

On kagom\'e lattice, 
similar to triangular case, Hermele et al calculated the kagom\'e DSL with staggered flux mean-field ansatz, with three gamma matrices as $\gamma_\nu=(\mu^3,\mu^2,-\mu^1)$, and we have for the PSG of Dirac fermions as
\begin{align}
T_1: \psi\rightarrow (i\tau^2)\psi\quad T_2: \psi\rightarrow (i\tau^3)\psi\quad R_y: \psi\rightarrow (i\mu^1)exp(\frac{i\pi}{2}\tau_{ry}) \psi\nonumber\\
C_6: \psi\rightarrow exp(\frac{i\pi}{3}\mu^3) exp(\frac{2\pi i}{3}\tau_R)\psi\quad \mathcal T:\psi\rightarrow (i\sigma^2)(i\mu^2)(-i\tau^2)\psi.
\end{align}
where
\begin{align}
\tau_{ry}=\frac{-1}{\sqrt{2}} (\tau^1+\tau^3)\quad \tau_R=\frac{1}{\sqrt{3}}(\tau^1+\tau^2-\tau^3).
\end{align}

\begin{table*}
\captionsetup{justification=raggedright}
\begin{center}
\begin{tabular}{|p{13mm}|c|c|c|c|c|c|}
\hline
Lattice&Bilinears& $T_1$ & $T_2$ & $ Reflection$ & $ Rotation$ & $\mathcal T$\\
\hline

\multirow{8}{*}{square}&$M_{00}$ &$+$&$+$&$-$&$+$&$-$\\
&$M_{i0}$ &$-$ &$-$& $-$& $-$& $-$ \\
&$M_{01}$ &$-$&$+$&$+$& $M_{03}$&$+$\\
&$M_{02}$ &$+$&$+$&$+$& $-M_{02}$&$-$\\
&$M_{03}$ & $+$ & $-$ &$-$& $-M_{01}$&$+$ \\

&$M_{i1}$ & $+$ & $-$ & $+$ &$-M_{i3}$& $+$\\
&$M_{i2}$ & $-$ & $-$ & $+$ & $M_{i2}$& $-$\\
&$M_{i3}$ & $-$ & $+$ & $-$ &$M_{i1}$& $+$\\
\hline

\multirow{8}{*}{\parbox{8mm}{honey-\\ comb}}&$M_{00}$ &$+$&$+$&$-$&$+$&$-$\\
&$M_{i0}$ & $+$ & $+$& $-$& $+$& $+$ \\
&$M_{01}$ & \multicolumn{2}{c|}{$ \cos(\frac{2\pi}{3}) M_{01}+\sin(\frac{2\pi}{3}) M_{02}$} & $+$ & $\cos(\frac{2\pi}{3}) M_{01}+\sin(\frac{2\pi}{3}) M_{02}$&$+$\\
&$M_{02}$ & \multicolumn{2}{c|}{$ \cos(\frac{2\pi}{3}) M_{02}-\sin(\frac{2\pi}{3}) M_{01}$} & $-$ & $-\cos(\frac{2\pi}{3}) M_{02}+\sin(\frac{2\pi}{3}) M_{01}$&$+$\\
&$M_{03}$ & $+$ & $+$ &$+$& $-$&$+$ \\

&$M_{i1}$ &  \multicolumn{2}{c|}{$ \cos(\frac{2\pi}{3}) M_{i1}+\sin(\frac{2\pi}{3}) M_{i2}$}  & $+$ &$\cos(\frac{2\pi}{3}) M_{i1}+\sin(\frac{2\pi}{3}) M_{i2}$ & $-$\\
&$M_{i2}$ & \multicolumn{2}{c|}{$ \cos(\frac{2\pi}{3}) M_{i2}-\sin(\frac{2\pi}{3}) M_{i1}$} & $-$ & $-\cos(\frac{2\pi}{3}) M_{i2}+\sin(\frac{2\pi}{3}) M_{i1}$ & $-$\\
&$M_{i3}$ & $+$ & $+$ & $+$ & $-$ & $-$\\
 \hline
\multirow{8}{*}{\parbox{8mm}{triangle}}&$M_{00}$ & $+$ &$+$& $-$&$+$& $-$\\

 &$M_{i0}$&$+$ &$+$& $+$&$-$ & $+$\\

 &$M_{01}$& $-$ & $-$ & $   -M_{03} $ &  $   -M_{02}  $ & $+$\\
  &$M_{02}$& $+$ & $-$ & $   M_{02} $ &  $  M_{03}  $  & $+$\\
 &$M_{03}$& $-$ & $+$ & $   -M_{01}  $ &  $   M_{01}  $ & $+$ \\

 &$M_{i1}$& $-$ & $-$ & $   M_{i3}$ &  $   M_{i2} $ & $-$ \\
&$M_{i2}$& $+$ & $-$ & $   -M_{i2} $ &  $  -M_{i3}$ & $-$ \\
&$M_{i3}$& $-$ & $+$ & $  M_{i1} $ &  $  -M_{i1} $ & $-$ \\
   \hline
\multirow{8}{*}{\parbox{8mm}{kagome}}&$M_{00}$ &$+$&$+$&$-$&$+$&$-$\\
&$M_{i0}$ & $+$ & $+$& $-$& $+$& $+$\\
&$M_{01}$ & $-$ & $-$ & $-M_{03}$ & $M_{02}$ & $+$ \\
&$M_{02}$ & $+$ & $-$ & $M_{02}$ & $-M_{03}$ & $+$\\
&$M_{03}$ & $-$ & $+$ &$-M_{01}$& $-M_{01}$&$+$\\

&$M_{i1}$ & $-$ & $-$ & $-M_{i3}$ & $M_{i2}$& $-$\\
&$M_{i2}$ & $+$ & $-$ &$M_{i2}$& $-M_{i3}$ & $-$\\
&$M_{i3}$ & $-$ & $+$ & $-M_{i1}$ & $-M_{i1}$ & $-$\\
\hline
\end{tabular}
\caption{The transformation of fermion bilinears $M_{ij}\equiv \overline \psi \sigma^i\tau^j\psi$ on four types of lattices(staggered flux mean field on square lattice). $T_{1/2}$, reflection are marked in fig~\ref{fig:honeycomb2}, rotation denotes $4-$fold rotation around site for square, $6-$fold rotation for honeycomb/kagome/triangular lattices. }\label{table:bilinears}
\end{center}
\end{table*}

\section{The $SO(3)_s\times SO(3)_v\times U(1)_{top}$ anomaly}
\label{SSUanomaly}

We take the QED$_3$ Lagrangian and gauge the $SO(3)_s\times SO(3)_v\times U(1)_{top}$, with gauge connections denoted by $\mathcal{A}_s, \mathcal{A}_v, A_{top}$, respectively. We also regularize the Dirac fermions using a Pauli-Villars regulator (which is different from the notation used in the main text), and define the Euclidean partition function of a Dirac fermion coupled with a gauge field $A$ and metric $g$ to be
\be
Z[A,g]_{PV}=|Z[A,g]|{\rm exp}\left(-\frac{i\pi}{2}\eta[A,g] \right),
\ee
where $\eta[A,g]$ is the $\eta$-invariant, which is similar to a half-level Chern-Simons term classically, but is gauge invariant unlike the half-level Chern-Simons term (see Ref.~\cite{wittenreview} for more details). One can simply interpret the $\eta$-invariant as coming from an additional gapped Dirac fermion. Now the gauged QED$_3$ theory should be properly written as
\be
S=\int\left[(\bar{\psi}\slashed{D}_{a/2,A_s,A_v,g}\psi)_{PV}\nn+\frac{i}{2}CS[a]+\frac{i}{2}CS[A_s]+\frac{i}{2}CS[A_v]+4CS[g]+\frac{i}{4\pi}adA_{top}\right],
\ee
where the Chern-Simons terms in $a, A_s, A_v, g$ are introduced to compensate the $\eta$-invariant at classical level (notice we have also redefined $a\to a/2$ for later convenience), and the last term comes from the definition of $U(1)_{top}$. However those terms are not strictly gauge invariant in $(2+1)d$. Instead we should consider a $4d$ (orientable) bulk system and view our $3d$ system as its boundary. The Chern-Simons terms can now be written as perfectly gauge invariant bulk $\Theta$-terms
\be
\label{bulkTheta}
S_{bulk}=\frac{i\pi}{2}\left(p_1[a]+p_1[A^s]+p_1[A^v]-\sigma[M_4]+2\int_{M_4}\frac{da}{2\pi}\wedge\frac{dA}{2\pi} \right),
\ee
where $p_1$ is the first Pontryagin number of the $SO(N)$ bundles (here $a$ is viewed as an $SO(2)$ bundle)
\be
p_1=\frac{1}{8\pi^2}\int_{M_4}tr(F\wedge F),
\ee
and $\sigma$ is the signature of the manifold
\be
\sigma=-\frac{1}{24\pi^2}\int_{M_4}tr(R\wedge R),
\ee
where $F$ and $R$ are the gauge field strength and Riemann tensor, respectively. Both $p_1$ and $\sigma$ are integers. We also have the relation\footnote{This holds when $w_4$ vanishes, which is true for $SO(2)$ and $SO(3)$ bundles. }
\be
\label{pontryaginsquare}
p_1=\mathcal{P}(w_2)\hspace{5pt} ({\rm mod}\hspace{2pt}4),
\ee
where $\mathcal{P}$ denotes Pontryagin square operation and $w_2$ denotes the second Stifel-Whitney class of the $SO(N)$ bundle. On any $2$-cycle we should have the constraint (cocycle condition)
\be
w_2^a +w_2^s+w_2^v+w_2^{TM}=0 \hspace{5pt}({\rm mod}\hspace{2pt} 2),
\ee
where $w_2^a,w_2^s,w_2^v,w_2^{TM}$ are the second Stifel-Whitney classes of the $a, SO(3)_s, SO(3)_v$ and tangent bundles, respectively. This condition simply comes from the fact that any field that carries half charge under $a$ (unit charge in the notation in main text) must also carry half-spin under $SO(3)_s$ and $SO(3)_v$, and must be a fermion. Putting the cocycle condition and Eq.~\eqref{pontryaginsquare} into the bulk Theta terms Eq.~\eqref{bulkTheta}, and using the facts that (a) $\mathcal{P}[a+b]=\mathcal{P}[a]+\mathcal{P}[b]+2a\cup b (mod 4)$, (b) $\mathcal{P}[a]=a\cup a (mod 2)$, (c) $a\cup w_2^{TM}=a\cup a$ for $a\in H^2(M,\mathbb{Z}_2)$, (d) $\mathcal{P}(w_2^{TM})=\sigma (mod 4)$, and (e) $w_2^a=da/2\pi (mod 2)$ for $SO(2)$ bundles, we have
\be
S_{bulk}=i\pi\int_{X_4}\left[ w_2^s\cup w_2^v+\left(w_2^s+w_2^v+\frac{dA_{top}}{2\pi} \right)\cup \frac{dA_{top}}{2\pi}  \right],
\ee
as promised in the main text. Notice that the bulk term only depends on the ``external" gauge fields and does not depend on $a$ -- this is required since $a$ exists as a dynamical gauge field defined only in the $3d$ field theory.

\end{widetext}

\bibliography{monopole_overleaf}

\end{document}